\documentclass[11pt,a4paper]{article} 
\usepackage{booktabs}
\usepackage{multirow}
\usepackage{jheppub}
\usepackage{mciteplus}
\usepackage{graphicx}
\usepackage{dcolumn}
\usepackage{mathrsfs}
\usepackage{slashed}
\usepackage{subfigure}
\usepackage{afterpage}

\newcommand{\Dmq}{\Delta m^2}

\def\dcp{\delta_{\rm CP}}
\def\oldcp{\overline{\dcp}}

\def\be{\begin{equation}}
\def\ee{\end{equation}}
\def\beq{\begin{equation}}
\def\eeq{\end{equation}}
\def\bc{\begin{center}}
\def\ec{\end{center}}
\def\bea{\begin{eqnarray}}
\def\eea{\end{eqnarray}}

\newcommand{\mean}[1]{\langle#1\rangle}

\newcommand{\mbD}{\mathbf{D}}
\newcommand{\mbTh}{\mathbf{\Theta}}

\newcommand{\lhood}{\mathcal{L}}
\newcommand{\ev}{\mathcal{Z}}

\newcommand{\mcB}{\mathcal{B}}
\newcommand{\AIC}{{\rm AIC}}

\newcommand{\ie}{\emph{i.e.}}
\newcommand{\eg}{\emph{e.g.}}
\newcommand{\df}{{\rm d}}
\newcommand{\im}{{\rm i}}
\newcommand{\qu}[1]{``#1''}

\def\T2K{{\sc T2K }}

\def\MN{{\sc MultiNest}}

\newcommand{\refcite}[1]{Ref.~\cite{#1}}
\newcommand{\figref}[1]{Fig.~\ref{#1}}
\newcommand{\tabref}[1]{Tab.~\ref{#1}}
\newcommand{\secref}[1]{Sec.~\ref{#1}}

\def\NO{{\rm NO}}
\def\IO{{\rm IO}}

\newcommand{\snq}[1]{s^{2}_{#1}}

\title{Bayesian global analysis of  neutrino oscillation data}
\author[a]{Johannes Bergstr\"om,}
\affiliation[a]{Departament d'Estructura i Constituents de la Mat\`eria and Institut  de Ciencies del Cosmos,\\ Universitat de Barcelona, Diagonal 647,  E-08028 Barcelona, Spain}

\author[a,b,c]{M.~C.~Gonzalez-Garcia,}
\affiliation[b]{Instituci\'o Catalana de Recerca i Estudis
  Avan\c{c}ats (ICREA)}
\affiliation[c]{C.N.~Yang Institute for Theoretical Physics, State
  University of New York at Stony Brook, Stony Brook, NY 11794-3840,
  USA}
\emailAdd{maria.gonzalez-garcia@stonybrook.edu}

\author[d]{Michele Maltoni,}
\affiliation[d]{Instituto de F\'{\i}sica Te\'orica UAM/CSIC, Calle de
  Nicol\'as Cabrera 13--15, Universidad Aut\'onoma de Madrid,
  Cantoblanco, E-28049 Madrid, Spain}
\emailAdd{michele.maltoni@csic.es}

\author[e]{Thomas Schwetz}
\affiliation[e]{Oskar Klein Centre for Cosmoparticle Physics,
  Department of Physics, Stockholm University, SE-10691 Stockholm,
  Sweden}
\emailAdd{schwetz@fysik.su.se}

\abstract{
We perform a Bayesian analysis of current neutrino oscillation
data. When estimating the oscillation parameters we find that the
results generally agree with those of the $\chi^2$ method,
with some differences involving $\snq{23}$ and CP-violating
effects. We discuss the additional subtleties caused by the circular
nature of the CP-violating phase, and how it is possible to obtain
correlation coefficients with $\snq{23}$. When performing model
comparison, we find that there is no significant
evidence for any mass ordering, any octant of $\snq{23}$ or a deviation from maximal mixing, nor 
the presence of CP-violation }

\preprint{IFT-UAM/CSIC-15-072, YITP-SB-15-24}

\begin{document}

\maketitle

\section{Introduction}
Neutrino oscillation experiments have now established beyond doubt
that neutrinos are massive and there is leptonic flavour violation
in their propagation 
~\cite{Pontecorvo:1967fh,Gribov:1968kq}, 
see Ref.~\cite{GonzalezGarcia:2007ib} for an overview. 
It has also been clear for more than a decade that 
a consistent description of the global data on neutrino
oscillations is possible by assuming that the three known
neutrinos ($\nu_e$, $\nu_\mu$, $\nu_\tau$) are linear
quantum superposition of three massive states $\nu_i$ ($i=1,2,3$)
with masses $m_i$. Consequently, a leptonic mixing matrix is present
in the weak charged current interactions~\cite{Maki:1962mu,
  Kobayashi:1973fv} of the mass eigenstates, 
which can be parametrized as~\cite{PDG}:
\begin{equation}
  \label{eq:matrix}
  U =
  \begin{pmatrix}
    c_{12} c_{13}
    & s_{12} c_{13}
    & s_{13} e^{-i\delta_\text{CP}}
    \\
    - s_{12} c_{23} - c_{12} s_{13} s_{23} e^{i\delta_\text{CP}}
    & \hphantom{+} c_{12} c_{23} - s_{12} s_{13} s_{23}
    e^{i\delta_\text{CP}}
    & c_{13} s_{23}
    \\
    \hphantom{+} s_{12} s_{23} - c_{12} s_{13} c_{23} e^{i\delta_\text{CP}}
    & - c_{12} s_{23} - s_{12} s_{13} c_{23} e^{i\delta_\text{CP}}
    & c_{13} c_{23}
  \end{pmatrix},
\end{equation}
where $c_{ij} \equiv \cos\theta_{ij}$ and $s_{ij} \equiv
\sin\theta_{ij}$.  If one chooses the convention where the angles 
$\theta_{ij}$ are taken to lie in the first quadrant, $\theta_{ij} \in [0, \pi/2]$,  and the CP phase $\delta_\text{CP} \in [0, 2\pi]$, 
then $\Delta m^2_{21}=m_2^2-m_1^2>0$ by convention, and  $\Delta m^2_{31}$
can be positive or negative.
It is customary to refer to the first option as Normal Ordering (NO), and
to the second one as Inverted Ordering (IO).
In the following we adopt the (arbitrary) convention of reporting
results for $\Dmq_{31}$ for NO and $\Dmq_{32}$ for IO, \textit{i.e.},
we always use the one which has the larger absolute value. Sometimes
we will generically denote such quantity as $\Dmq_{3\ell}$, with
$\ell=1$ for NO and $\ell=2$ for IO.

Several global analyses exist in the literature 
\cite{Gonzalez-Garcia:2014bfa,Capozzi:2013csa,Forero:2014bxa},
which, 
by fitting the results from the bulk of oscillation experiments, obtain
best estimates and allowed ranges for these six oscillation parameters. 
Generically they obtain their results within a frequentist framework, using 
a $\chi^2$ statistics.

Alternatively, a consistent approach to obtaining the probability that a certain
parameter within a given model takes certain values is provided by 
Bayesian inference.  Furthermore,  
Bayesian analysis is particularly
suited for comparing how much better one model describes the
data compared to another model. 
So one may question to what degree the current determination of the 
oscillation parameters is dependent on the assumed statistical approach,
and whether Bayesian statistics can shed some light on the presently 
open issues related to the mass ordering, the octant of $\theta_{23}$,
and the presence of CP-violation. 
 
In this article we address these questions by performing a Bayesian analysis of
the current neutrino oscillation data. In Sec.~\ref{sec:introstat} we
briefly describe the elements of Bayesian statistics required for this
analysis. In Sec.~\ref{sec:posterior} we  present the global results of
the analysis and compare them with those of the $\chi^2$ analysis 
of the same data samples of NuFIT 2.0~\cite{nufit}. 
We discuss in detail the main results
related to the determination of $\sin^2\theta_{23}$ and $\delta_{\rm CP}$
in Secs.~\ref{sec:theta23} and~\ref{sec:deltacp}, where we also 
discuss the additional subtleties caused by the circular
nature of the CP-violating phase, and study how it is possible to define
correlation coefficients with $\snq{23}$
in Sec.~\ref{sec:corre}.  
Finally in \secref{sec:conclusions} we summarize our conclusions.

\section{Statistical framework}
\label{sec:introstat}
In this work, we will be using Bayesian probability theory, where each
proposition is associated with a probability or \emph{plausibility},
defined to lie between 0 and 1. In order to calculate the
probabilities of different assumptions, hypotheses, or models, the
laws of probability are used when conditioned on some known (or
assumed) information. Of particular interest is \emph{Bayes' theorem},
which can be used to compare a set of hypotheses $M_j$, using some set
of collected data, $\mbD$, through calculation of the \emph{posterior
  odds},
\begin{equation}\label{eq:post_ratio} 
\frac{ \Pr(M_i|\mathbf{D})}{\Pr(M_j|\mathbf{D})} =
\frac{\Pr(\mathbf{D}|M_i)}{\Pr(\mathbf{D}|M_j)}
\frac{\Pr(M_i)}{\Pr(M_j)}.
\end{equation}
The \emph{prior odds} $\Pr(M_i)/ \Pr(M_j)$ quantifies how much more
plausible one model is than the other \emph{a priori}. The
\emph{evidence}, $\ev_i = \Pr(\mathbf{D}|M_i)$, is the likelihood for
the model quantifying how well the model describes the data. The
\emph{Bayes factor}, 
\beq
\mcB = \ev_i/\ev_j
\eeq  
which is the ratio of the
evidences, quantifies how much better the model $M_i$ describes the
data than $M_j$.

Given that the model $M$ contains the free parameters $\mbTh$, the evidence is given by
\bea
\mathcal{Z} =\Pr(\mathbf{D}|M) &=& 
\int{\mathcal{L}(\mathbf{\Theta})\pi(\mathbf{\Theta})}\df^N\mathbf{\Theta},
\label{eq:Z}
\eea
where $\mathcal{L}(\mathbf{\Theta}) \equiv
\Pr(\mathbf{D}|\mathbf{\Theta}, M)$ is the \emph{likelihood function}.
The prior probability density of the parameters is given by
$\pi(\mathbf{\Theta}) \equiv \Pr(\mathbf{\Theta}|M)$, and should
always be normalized, \ie, it should integrate to unity. The
assignment of priors are probably the most discussed and controversial
part of Bayesian inference. This is often far from trivial, but
nevertheless this assignment is an important, even essential, part of any Bayesian analysis.

The Bayes factors, or rather the posterior odds, are interpreted or
\qu{translated} into ordinary language using the so-called
\emph{Jeffreys scale}, given in \tabref{tab:Jeffreys} as used in, \eg,
Refs.~\cite{Trotta:2008qt,Hobson:2010book} (\qu{$\log$} denotes the
natural logarithm). Even though the Bayes factor in general will
favour the correct model once \qu{enough} data have been obtained, the
evidence is often highly dependent on the choice of prior on the parameters.

\begin{table}
\begin{center}
\begin{tabular}{@{}llll@{}}
\hline
$|\log(\text{odds})|$ & odds & $\Pr(M_1 | \mathbf{D})$ & Strength of evidence\\ 
\hline
$<1.0$ & $\lesssim 3:1$ & $\lesssim 0.75$ & Inconclusive \\
$1.0$ & $\simeq 3:1$ &  $\simeq 0.75$ & Weak evidence \\
$2.5$ & $\simeq 12:1$ & $\simeq 0.92$ & Moderate evidence \\
$5.0$ & $\simeq 150:1$ & $ \simeq 0.993$ & Strong evidence \\
\hline
\end{tabular}
\end{center}
\caption{\it The Jeffreys scale, used for interpretation of Bayes factors, odds, and model probabilities. The posterior model probabilities for the preferred model are calculated assuming only two competing hypotheses and equal prior probabilities. Note that $\log$ denotes the natural logarithm.}
\label{tab:Jeffreys}
\end{table}

In principle, the evidence defined above is really the only consistent 
quantity to judge the (relative) merit of a model. However, 
there are also some so-called
\emph{information criteria} which have been used to compare different
models, see, \eg, \cite{Burnham:2003,Liddle:2007}. These do not
explicitly depend on any prior, but typically are derived using quite
restrictive assumptions. This makes their use less reliable, since
conclusions based on them could differ much from a full Bayesian
analysis. We will also consider the Akaike Information Criterion (AIC)
(which is neither a Bayesian nor a frequentist meassure),
motivated by minimizing the expected \qu{distance} between the true
data distribution, and the data distribution given by the fitted
model. It yields a fixed penalty to each model as\footnote{The factor
  of 2 is just for historical reasons. There is also a modified
  criterion for small sample sizes, which we do not consider here
  since the number of samples is rather large.}  \be \AIC = - 2 \log
\lhood_{\rm max} + 2 N_{\rm par} = \chi^2_{\rm min} + 2 N_{\rm par}
,\ee dropping an irrelevant constant, and with $N_{\rm par}$ the
number of free parameters. Hence, we see that each additional parameter needs to improve
the $\chi^2$ by 2 units to make up for the additional complexity. Although great
caution should be exercised, typically $\widetilde{\ev} \propto
e^{-\AIC/2} = \lhood_{\rm max} e^{- N_{\rm par}} $ would be used as a
proxy for the model likelihood, and hence $-\Delta \AIC/2$ between two
models as log of the Bayes factor, and interpreted using
\tabref{tab:Jeffreys}. However, unlike the Bayesian evidence, it
punishes complex models with additional parameters regardless of
whether these are constrained by the data, and for parameters which
are constrained, the punishment is typically smaller than in the full
Bayesian analysis.


Under the assumption that a model $M$ is true, complete inference of its parameters is given by the posterior distribution,
\begin{equation} 
\label{eq:Bayes_params} 
\Pr( \mathbf{ \Theta} | \mathbf{D},M) = 
\frac{\Pr(\mathbf{D}
|\mathbf{\Theta},M)\Pr(\mathbf{\Theta}|M)}
{\Pr(\mathbf{D}|M)}  = \frac{\lhood(\mbTh)\pi(\mbTh)}{\ev}.
\end{equation}
In this case, the evidence is only a normalization factor, since it is
independent of the values of the parameters $\mbTh$ and it is
therefore often disregarded in parameter estimation. 
Thus the main result of Bayesian parameter inference is the posterior and
its marginalized versions (usually in one or two
dimensions).  In this respect, one must distinguish between the 
marginal posterior
distributions and the marginal likelihood, which is the likelihood
integrated over all other parameters (after multiplication by the
prior of these parameters). The former is a probability distribution,
while the latter is not \cite{berger1999}. However, if the parameters
of interest have a uniform prior, the marginal posterior distribution
and the marginal likelihood are proportional to each other. For the
present analysis, it is only for the derived parameter $J_{\rm CP}$
that the prior is sufficiently non-uniform to have a noticeable impact
on the posterior, as we will show in Sec.~\ref{sec:deltacp}.

Generically in parameter inference, 
point estimates such as the posterior mean or
median are given together with \emph{credible intervals (regions)}
for the parameters.
A common way to define Bayesian credible intervals for a given parameter 
is by including all 
values with a posterior above a certain value, which however makes them
non-invariant under non-linear reparametrizations.  Invariance can be
restored by defining them to be iso-marginal likelihood intervals
instead.
\footnote{Although this only makes sense, as is the case here,
  with a clear separation of data and prior information, the latter
  being negligible.} Then, one calls the \qu{credible level} of a
value $\eta = \eta_0$ of a subset of parameters simply the posterior
volume within the likelihood of that value, 
\be {\rm CL}(\eta_0) =
\int_{\lhood(\eta) > \lhood(\eta_0) } \Pr (\eta | \mbD ) \df \eta.
\label{eq:CL}
\ee
This function is converted to the \qu{number of $\sigma$'s} in
the usual manner as
\be 
S = \sqrt{2} {\rm erfc}^{-1}(1-{\rm CL}). 
\label{eq:S}
\ee

In this work we use
\MN~\cite{Feroz:2007kg,Feroz:2008xx,Feroz:2013hea}, a Bayesian
inference tool which, given the prior and the likelihood, calculates
the evidence with an uncertainty estimate, and generates posterior
samples from distributions that may contain multiple modes and
pronounced (curving) degeneracies in high dimensions.

\subsection{Priors on oscillation parameters}
In a Bayesian analysis one has to choose a prior on model parameters,
in our case the mixing parameters and mass-squared differences. Before
considering any data, this prior should preferably not favour any
basis or direction in flavour space, \ie, be invariant under
rotations, or group transformations \cite{Berger:1985}. This Haar
measure of neutrino mixing matrices is, after integrating out
nonphysical and potential Majorana phases, the separable measure
\cite{Haba:2000be} \be \pi(s^2_{12}, c^4_{13}, s^2_{23}, \dcp) =
1/360^\circ, \ee in the standard parameterization. Although the prior
is uniform in $c^4_{13}$ and not, for example, $s^2_{13}$, this is of
no practical consequence since $s^2_{13}$ is well-measured and
significantly non-zero \refcite{Gonzalez-Garcia:2014bfa}. Furthermore,
using other, non-invariant, priors such as uniform in the angles will
in general not affect the results significantly.  On the mass-square
differences logarithmic priors are used. Since these are also
well-measured their prior is also of no practical significance.

In addition, the neutrino mass ordering can be considered as just
another free parameter.  In this way, the two orderings can be compared, and 
also the inference of other quantities can be performed not assuming a mass
ordering to be correct, but averaging over the two orderings. In this
last case we take $\pi(\NO) = \pi(\IO) = 0.5$,
and we denote this by \emph{mixed ordering} (MO).

Regarding the experimental nuisance parameters, they are all minimized
over as in a $\chi^2$ analysis. Since the uncertainties of these are
rather small  and Gaussian, including them in the Monte Carlo
and integrating over them instead of minimizing over them 
-- as would be the correct procedure in a
fully Bayesian analysis -- would make a negligible difference.

\section{Posterior distributions}
\label{sec:posterior}
First, under the assumption that three-neutrino mixing is
the true model, we perform parameter estimation and calculate 
the posterior  distributions of the six free parameters.  
In doing so we include the data from solar~\cite{Cleveland:1998nv,Kaether:2010ag,Abdurashitov:2009tn,Hosaka:2005um,
Cravens:2008aa,Abe:2010hy,sksol:nu2014,Aharmim:2011vm,Bellini:2011rx,Bellini:2008mr},
atmospheric~\cite{skatm:nu2014},
reactor~\cite{Gando:2010aa,Apollonio:1999ae,Piepke:2002ju,Abe:2012tg,db:nu2014,reno:nu2014,Declais:1994ma,Kuvshinnikov:1990ry,Declais:1994su,
Vidyakin:1987ue,Vidyakin:1994ut,Kwon:1981ua,Zacek:1986cu,Greenwood:1996pb,Afonin:1988gx}, and long baseline accelerator experiments
\cite{Adamson:2013whj,Adamson:2013ue,Abe:2014ugx,Abe:2013hdq}, in the
same data samples listed in Appendix of
Ref.~\cite{Gonzalez-Garcia:2014bfa} and used in NuFIT 2.0~\cite{nufit}. 

\begin{figure}[t!]
\begin{center}
\includegraphics[width=1\textwidth]{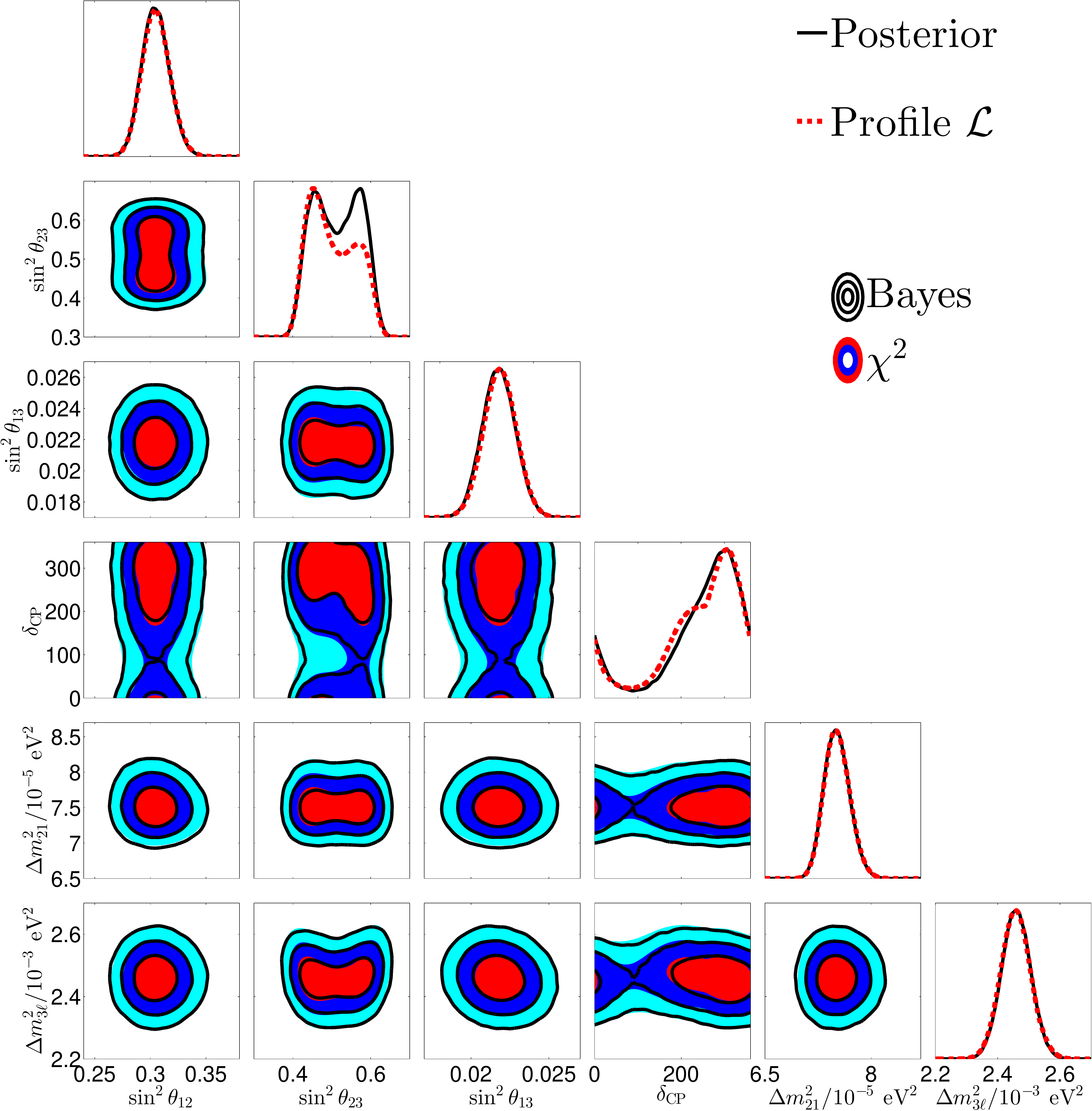}
\end{center}
\caption{One-dimensional posterior distributions (black full lines) 
and two-dimensional  $1\sigma$, $2\sigma$ and $3\sigma$ Bayesian credible 
regions  (black void contours). The figure also shows the one-dimensional 
profile likelihoods (red dashed curves)  and two-dimensional $\chi^2$ regions 
(coloured filled regions) from \refcite{Gonzalez-Garcia:2014bfa}.}
\label{fig:param_NO}
\end{figure}

\begin{figure}[t!]
\begin{center}
\includegraphics[width=1\textwidth]{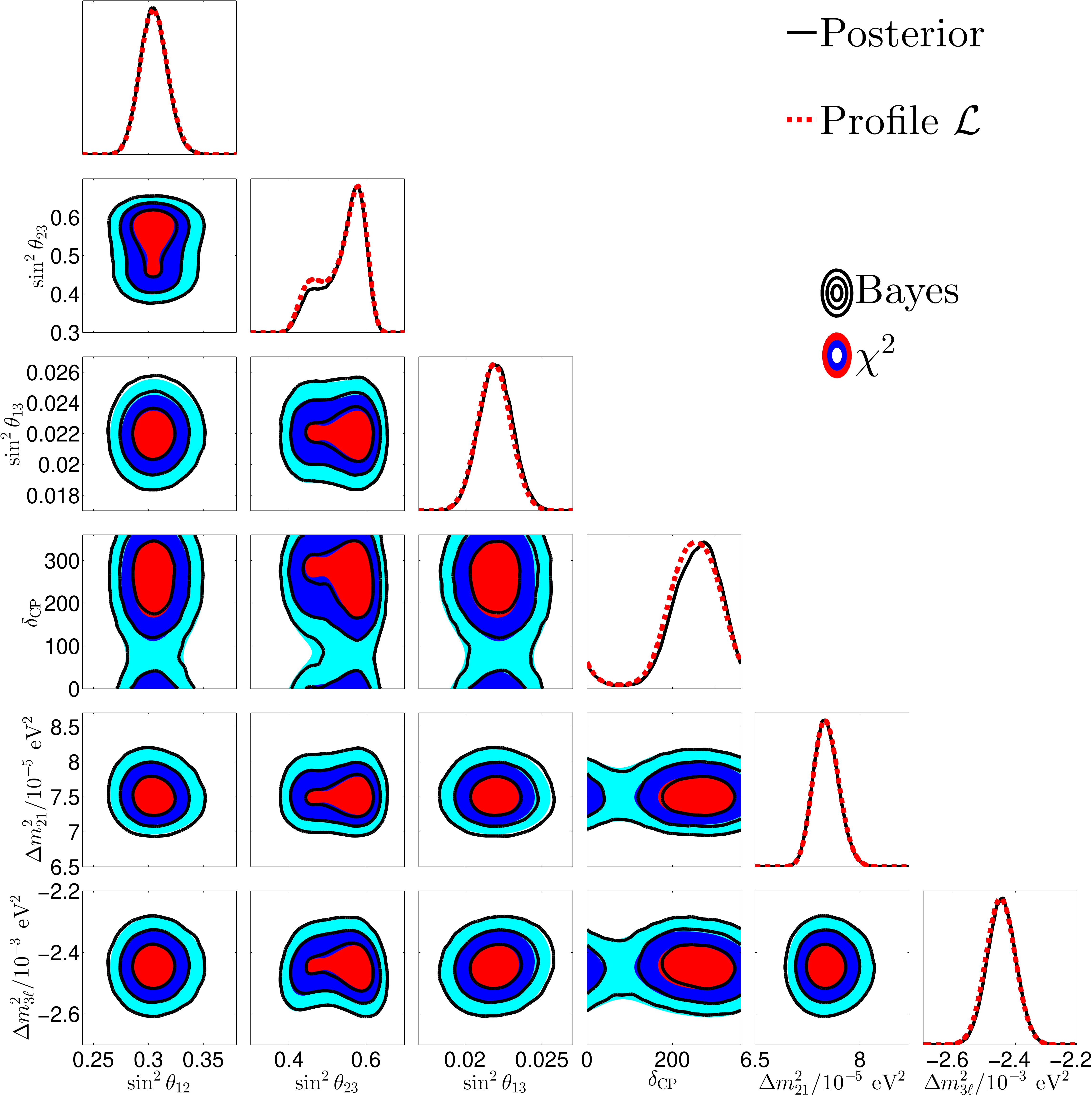}
\end{center}
\caption{Same as \figref{fig:param_NO} but for
  IO.}\label{fig:param_IO}
\end{figure}

\begin{figure}[t!]
\begin{center}
\includegraphics[width=1\textwidth]{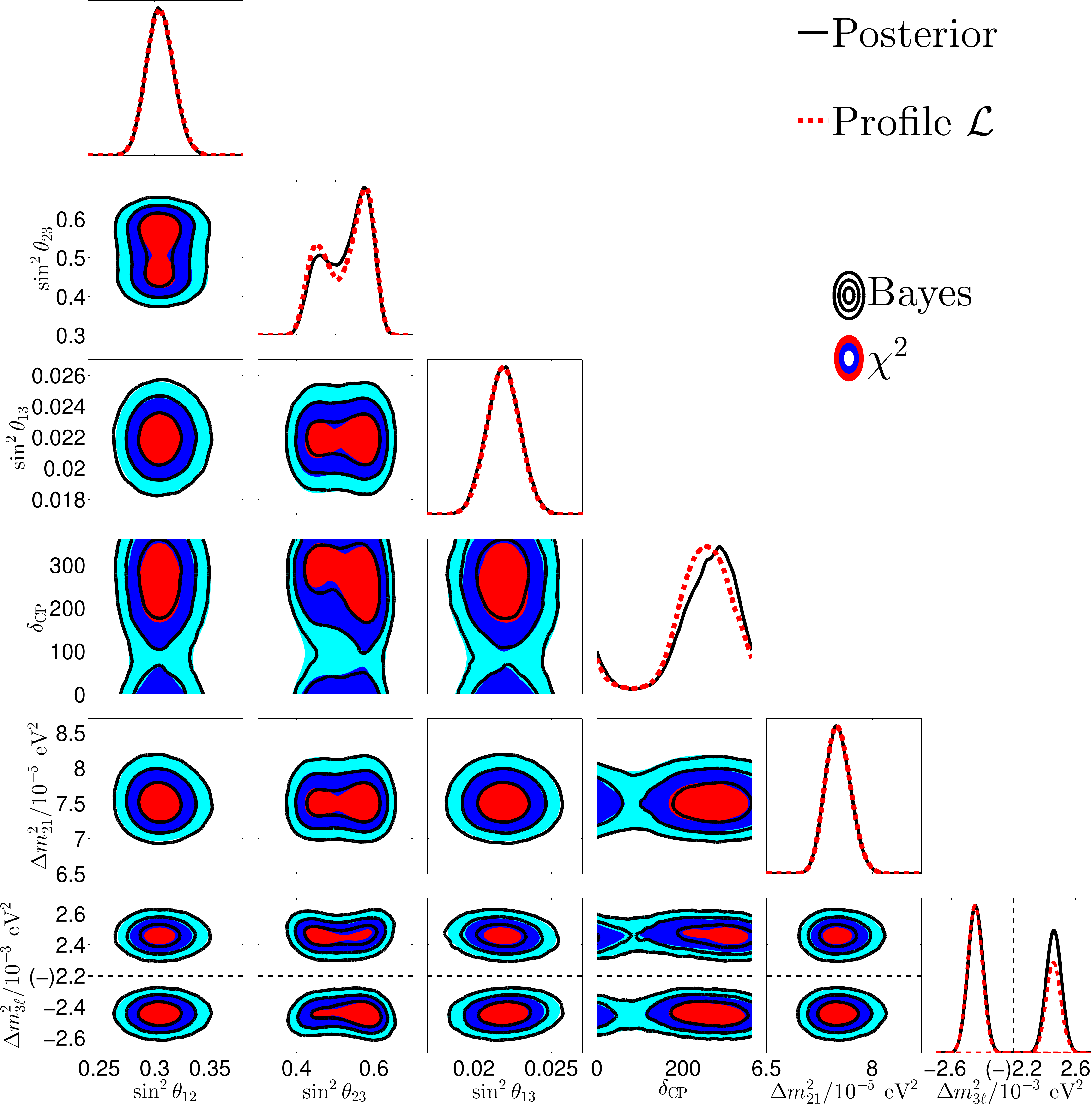}
\end{center}
\caption{Same as \figref{fig:param_NO} but for
  MO.}\label{fig:param_MO}
\end{figure}

The results are shown in  
\figref{fig:param_NO} for
NO, \figref{fig:param_IO} for IO, and \figref{fig:param_MO} for MO.
The posterior distribution for MO is simply the average of the NO and
IO posteriors, weighted by the posterior probabilities of the
orderings, 
\be 
\Pr(\mbTh | \mbD, {\rm MO}) = \sum_{O = {\rm NO,IO}}
\Pr(\mbTh | \mbD, O) \Pr(O | \mbD).  
\ee

From these figures, we conclude that the absolute values of the 
two mass-square differences, as well as the mixing angles, $\snq{12}$, and 
$\snq{13}$, are  well-measured and the posteriors of these parameters are
Gaussian to a very good approximation. 

We list in \tabref{tab:paramdet}  
different point estimates for each of these parameters:
the global maximum likelihood (which is the best fit point, bfp, of
the $\chi^2$ analysis), the point
at which the marginal likelihood is maximal, and the posterior mean and
median. The table also contains measures of the uncertainty of each
parameter in the form of the 1$\sigma$ and 3$\sigma$ 
Bayesian credible intervals as well as the corresponding
$\chi^2$   \emph{allowed regions} at the same CL (which we also 
call $\chi^2$ intervals for simplicity) 
which are identical to those given in \refcite{Gonzalez-Garcia:2014bfa}.  
As seen in the table, for these four parameters their Bayesian 
point estimates and uncertainties are practically indistinguishable from their 
$\chi^2$ counterparts.
Thus  we conclude  that the present determination of these four
parameters is very robust under variations of the statistical 
analysis and prior assumptions.

\begin{table}\centering
  \begin{footnotesize}
   \begin{tabular}
{l|@{\hskip 0.03cm}c@{\hskip 0.04cm}c@{\hskip 0.04cm}c@{\hskip 0.04cm}
c|c@{\hskip 0.04cm}c|c@{\hskip 0.04cm}c}
      \hline\hline
       \multicolumn{9}{c}{Normal Ordering}\\
\hline
 & 
\multicolumn{4}{c|}{Point Estimates} & 
\multicolumn{2}{c|}{ $\chi^2$ Intervals} & 
\multicolumn{2}{c}{ Bayes Credible Intervals}  \\
 & bfp & 
$\begin{matrix} {\rm max\; of}\\[-0.2cm]
 {\cal L}_{\rm marg}
\end{matrix}$ &mean & median &
1$\sigma$ CI & 3$\sigma$ CI & 1$\sigma$ CI & 3$\sigma$ CI \\\hline
$\sin^2\theta_{12}$ & 0.304 &  0.304  & 0.305   &  0.305   & 
[0.292,0.317]  & [0.270,0.344] & [0.292,0.317] 
                &          [0.269, 0.344]
\\ 
$\sin^2\theta_{13}$ & 0.0218 &  0.0218  & 0.0218  & 0.0218   & 
[0.0208,0.0228]  & [0.0186,0.0250] & [0.0207,0.0228] 
                &     [0.0187,0.0250]        
\\
$\frac{\Delta m^2_{21}}{10^{-5}{\rm eV}^2}$  & 7.5 &  7.5  & 7.5  &  7.5  & 
[7.33,7.69]  & [7.02,8.07] & [7.33,7.69]
                &  [7.03,8.09]            
\\
$\frac{\Delta m^2_{3\ell}}{10^{-3}{\rm eV}^2}$  & 2.457 &  2.460  &  2.459 & 2.459   & 
[2.417,2.504]  & [2.317,2.607] & [2.414,2.506]
                &   [2.320,2.601]          
\\
      \hline\hline
       \multicolumn{9}{c}{Inverted Ordering }\\
\hline
$\sin^2\theta_{12}$ & 0.304 &  0.305  &  0.305 &  0.305  & 
[0.292,0.317]  & [0.270,0.344] & [0.292,0.317]
                &     [0.269,0.344]        
\\ 
$\sin^2\theta_{13}$ & 0.0219 &   0.0219  &   0.0220 &  0.0220   & 
[0.0209,0.0230]  & [0.0188,0.0251] & [0.0209,0.0231] 
                &   [0.0189,0.0252]          
\\
$\frac{\Delta m^2_{21}}{10^{-5}{\rm eV}^2}$ & 7.5 &  7.5  &  7.5 &  7.5  & 
[7.33,7.69]  & [7.02,8.07] & [7.33,7.68] 
                &       [7.02,8.09]      
\\
$\frac{\Delta m^2_{3\ell}}{10^{-3}{\rm eV}^2}$ &- 2.449 &  - 2.445  & - 2.445  &  - 2.445  & 
[-2.496,-2.401]  & [-2.590,-2.307] & [-2.492,-2.400]  
                &    [-2.584,-2.308]          
\\
\hline\hline
    \end{tabular}
  \end{footnotesize}
  \caption{Comparison of the results of $\chi^2$  and Bayesian
 analysis in the framework of  three-flavor oscillations.
For comparison of the determination
of $\theta_{23}$ and $\delta_{\rm CP}$ see Sec.~\ref{sec:theta23}
and ~\ref{sec:deltacp}.}
  \label{tab:paramdet}
\end{table}

Considering the comparison between mass orderings,
we find that, assuming the same prior probability for both,  
their posterior probabilities are also very similar, the posterior probability of IO in this case given by
\begin{eqnarray}
\frac{
\Pr(\mathbf{D}|\rm IO)}
{\Pr(\mathbf{D}|\rm IO)+\Pr(\mathbf{D}|\rm NO)}=
\frac{\ev_{\rm IO}}{\ev_{\rm IO}+\ev_{\rm NO}} = 0.55.
\end{eqnarray}
The Bayes factor 
(which is independent of the prior on the ordering) is:
\be 
\log \mcB = \log \frac{\ev_\NO}{\ev_\IO} =\log\left(\frac{0.45}{0.55}\right)
=-0.2,  
\ee
\ie, there is a non-meaningful preference for inverted ordering. 
For comparison, the $\chi^2$ analysis finds $\Delta \chi^2 = 
\chi^2_{\rm min}(\rm{NO}) - \chi^2_{\rm min}(\rm{IO}) \simeq  0.97$. 
Trivially, this gives $\Delta \AIC/2 = 0.5$ in favor of IO, 
which is also what $\log \mcB$ would be if the likelihoods would have 
identical shapes. In summary, both $\Delta\chi^2$  and the Bayesian
model comparison agree that there is no evidence 
for any of the mass ordering in the present data. 
However one must not forget that since the mass ordering is 
not a continuous parameter, $\Delta \chi^2 $ should not have a 
$\chi^2$ distribution, and hence the quantification of the degree of 
favouring/disfavouring of a given ordering based on the corresponding
$\Delta\chi^2$ is not fully justified
(see \refcite{Blennow:2013oma} for further discussion). 


Finally we notice that figures \ref{fig:param_NO}--\ref{fig:param_MO} show 
some differences between the results of the $\chi^2$ and Bayesian analyses  
where $\dcp$ or $\snq{23}$ are
involved. For example, we see that the marginalization over $\dcp$
pulls the bulk of the posterior of $\snq{23}$ more into the second
octant. Motivated by these differences we present a more detailed 
study of the results on $\snq{23}$, $\dcp$, and CP-violation 
in the following sections.

\section{Determination of $\snq{23}$}
\label{sec:theta23}
\begin{figure}[h!]
\begin{center}
\includegraphics[width=0.49\textwidth]{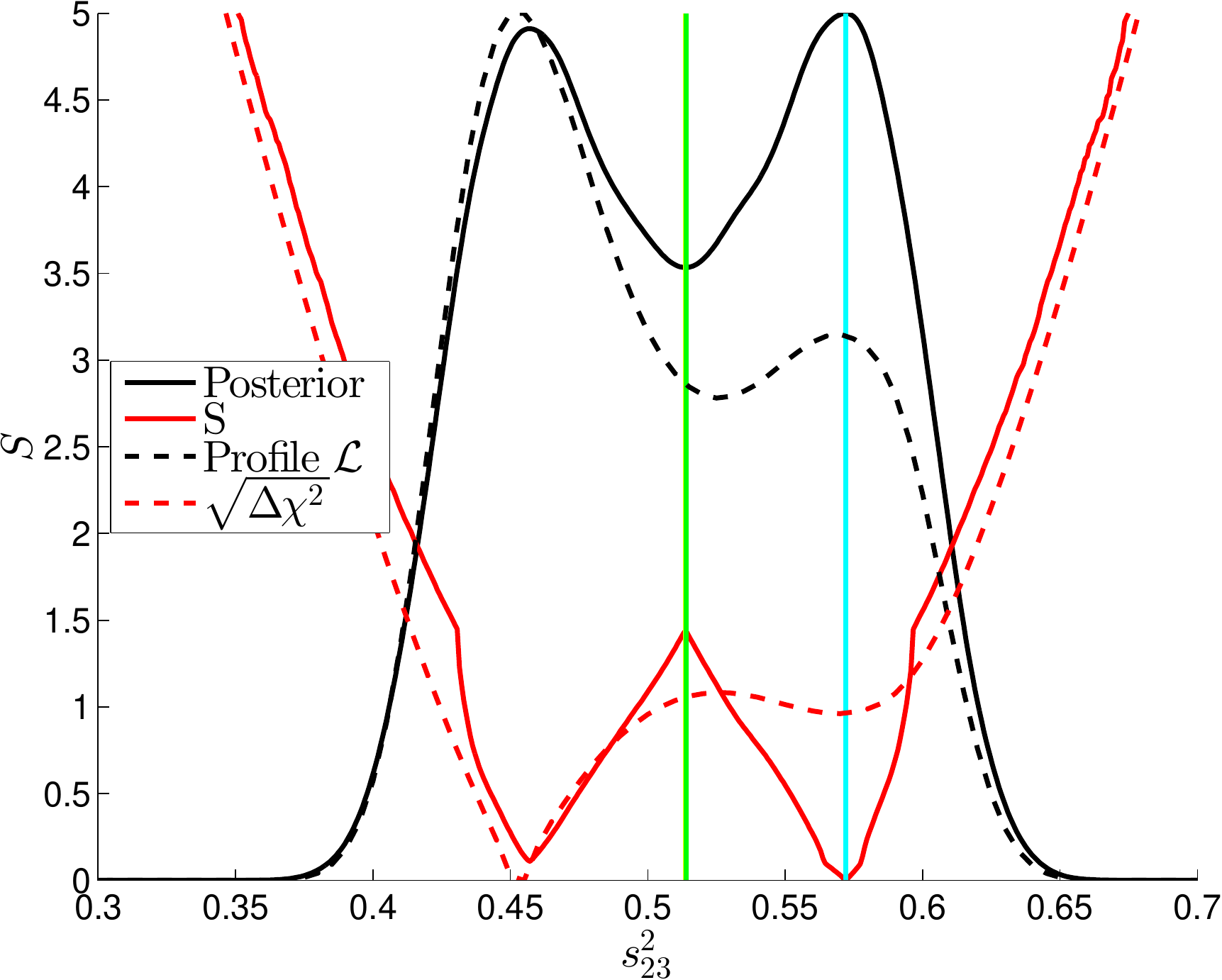}
\includegraphics[width=0.49\textwidth]{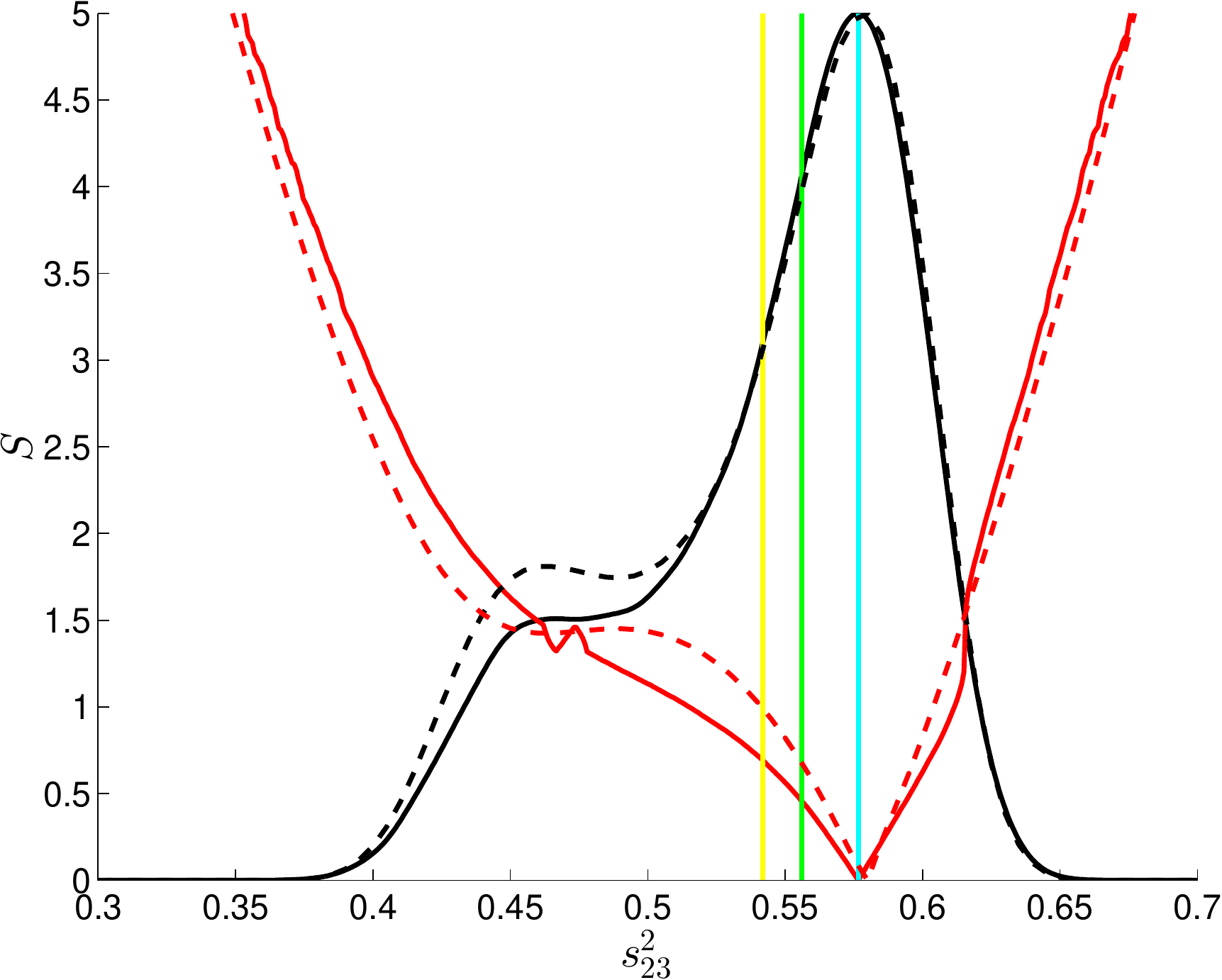}
\includegraphics[width=0.49\textwidth]{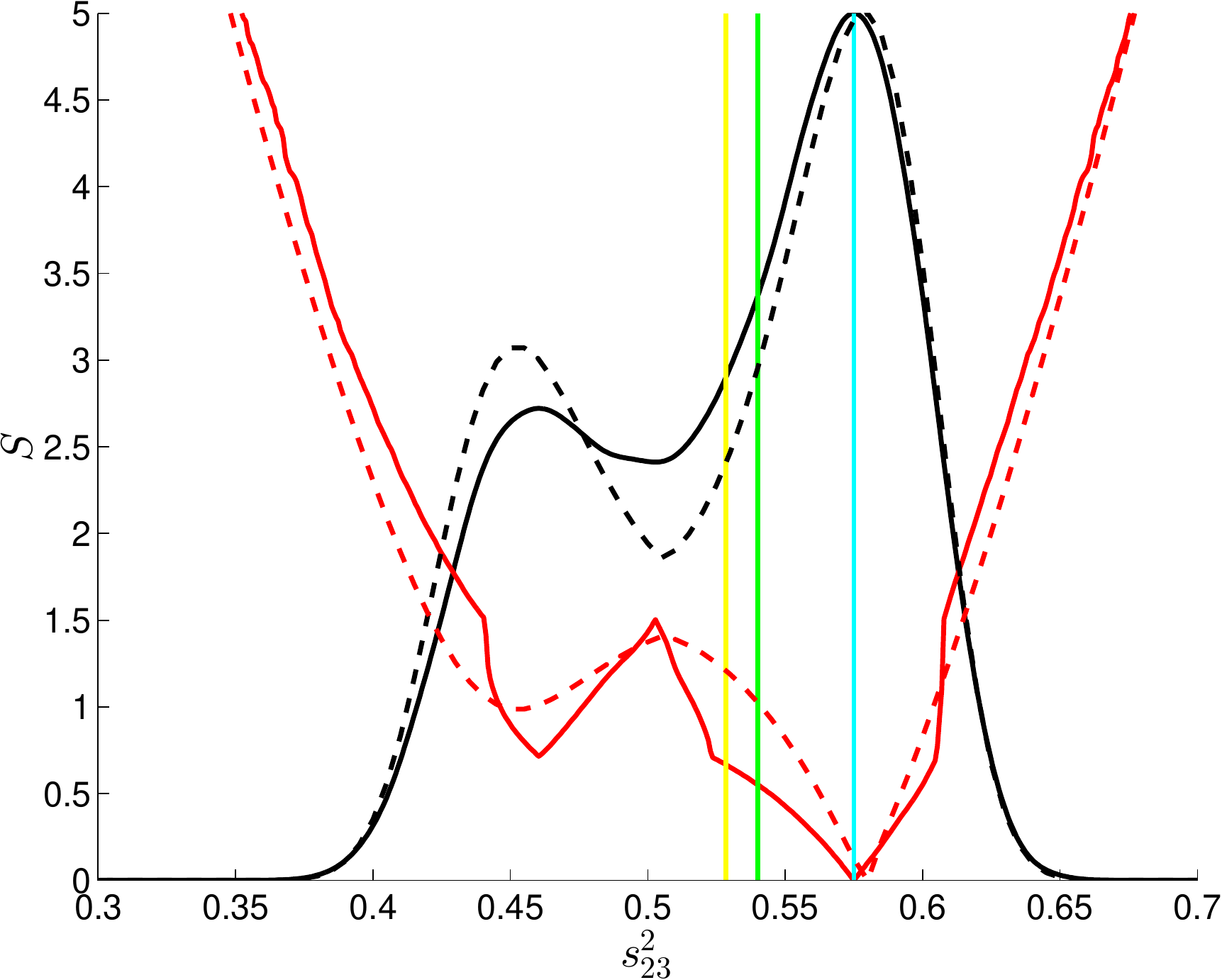}
\end{center}
\caption{Bayesian posterior/marginal likelihood (black solid), plotted
  together with the profile likelihood (black dashed), from
  \refcite{Gonzalez-Garcia:2014bfa} (both normalized to their maximal
  value). The number number of $\sigma's$)(red solid), and
  $\sqrt{\Delta \chi^2}$ (red dashed). Posterior mean (yellow line),
  median (green), and maximum of the marginal likelihood (cyan). NO
  (top left), IO (top right), MO (bottom).  }\label{fig:snq23_comb}
\end{figure}

In this section we study the determination of  $\snq{23}$ in more
detail. To do so, in \figref{fig:snq23_comb} 
we plot the Bayesian marginal posterior distribution 
(which in this case is proportional to 
the marginal likelihood) of $\snq{23}$ for all orderings 
together with the $S$ of the credible intervals 
(see Eqs.~(\ref{eq:CL}) and (\ref{eq:S})),  as well as the profile
likelihood and $\sqrt{\Delta \chi^2}$ (the nominal significance under
the assumption of a standard $\chi^2$ distribution).

We note that the Bayesian analysis generally prefers the second
octant and it does so more than the $\chi^2$ analysis, in particular for
NO. Although the credible and confidence levels differ in the vicinity
of the two peaks, both peaks are within the $2 \sigma$ region, and
outside of that region the difference between the two analyses is
rather small. Typically, the low-credibility Bayesian regions are larger than
the small-$\chi^2$ regions, while the high-credibility Bayesian regions are
smaller than the large-$\chi^2$ ones. This is just what is expected if
the likelihood contains a relatively sharp peak on top of a broader
plateau containing significant posterior probability.

For completeness, in addition to being displayed in 
\figref{fig:snq23_comb}, we also give the point estimates of $\snq{23}$ in
\tabref{tab:snq23_pointest}, namely, 
the global maximum likelihood,
the maximum of the marginal likelihood, and the posterior mean and
median. In \tabref{tab:snq23_unc} the 
measures of uncertainty are given in
the form of the posterior standard deviation, as well as credible
intervals corresponding to \figref{fig:snq23_comb}, and the regular
$\chi^2$ intervals.  

\begin{table}[h]
\begin{center}
\begin{tabular}{@{}rllll@{}} 
\hline
	Ordering	&	  Global max & max of $\lhood_{\rm marg}$ & mean  & median\\ 
\hline
 NO		& $0.452$ 	 &  $0.571$ 			& $0.515$& $0.516$ \\ 
 IO &	$0.579$ 			&  $0.576 $ 		& $0.541$& $0.555$ \\ 
MO 	&	$0.579$ 	& $0.576$  			& $0.529$& $0.542$  \\ \hline 
\end{tabular}
\caption{\it Point estimates of $\snq{23}$. \label{tab:snq23_pointest}}
\end{center}
\end{table}

\begin{table}[h]
\begin{center}
\begin{tabular}{@{}llllll@{}}
\hline
		Ordering	&	Method 	& st.~dev.		& $1 \sigma$ CI 				& $ 2 \sigma$ CI	& $ 3 \sigma$ CI \\ 
\hline
\multirow{2}*{NO}  	& Bayes	& $0.0585$  			& $[0.433,0.496],[0.530,0.594]$	&$[0.415,0.613]$	& $[0.389,0.637]$\\ 
	& $\chi^2$& -			 			& $[0.424,0.505],[0.554,0.582]$	&$[0.402,0.622]$	& $[0.381,0.643]$\\ 
\hline 
\multirow{2}*{IO} & Bayes		& $0.0534$ 				& $[0.514,0.612]$				& $[0.429,0.622]$	& $[0.400,0.640]$ \\ 
&$\chi^2$& -		 				& $[0.541,0.604]$				& $[0.416,0.625]$	& $[0.388,0.644]$ \\ 
\hline 
\multirow{2}*{MO}  &	Bayes	 	& $0.0574$ 				& $[0.449,0.476],[0.516,0.607]$	& $[0.422,0.618]$	& $[0.393,0.638]$  \\
&$\chi^2$& - 					& $[0.448,0.458],[0.541,0.604]$	& $[0.407,0.625]$	& $[0.385,0.644]$ 
\\ \hline 
\end{tabular}
\end{center}
\caption{\it Standard deviations, credible intervals, and
$\chi^2$ intervals for $\snq{23}$. \label{tab:snq23_unc}}
\end{table}

\subsection{Octants of $\theta_{23}$ and maximal mixing}
A related question  is that of which octant
$\theta_{23}$ belongs to, \ie, whether $s_{23}^2$ is larger or smaller
than $0.5$. With some similarity to the comparison of mass orderings,
this is also a comparison of two non-nested models with the same
number of parameters (although they are \qu{adjacent}), and so one cannot expect
difference between the $\chi^2$ minima between the two octants to have a
$\chi^2$ distribution. In a Bayesian analysis, the comparison is
however straightforward, by simply integrating the likelihoods over
each of the octants.

In addition, one can also consider maximal mixing, $s_{23}^2 = 0.5$, as a
realistic model, either exactly or approximately.  From a statistical
viewpoint, a model with a fixed value of a parameter can also be
interpreted as a model where there is some non-zero, but negligible
(compared to any experimental sensitivity) deviation from the fixed
value~\cite{hoeting1999}. Using any
of these viewpoints, i.e., by either considering exact maximal mixing
as a possible scenario, or alternatively as simply a very good
approximation, one can make a comparison with the octants.

As always, a model with additional parameters will be punished for
this extra complexity. In the present case, this punishment is
uniquely fixed by the compactness of the space of the allowed values
of $\snq{23}$. The Bayes factors between the second and first octants,
as well as between non-maximal and maximal mixing, are given in
\tabref{tab:MCs23}.\footnote{\refcite{Abe:2015awa} also compares the
octants and finds $\log \mcB = 0.6$ for all orderings for T2K data,
and $\log \mcB = 1.0 - 1.1$ when also including reactor data.}
\begin{table}[h]
\begin{center}
\begin{tabular}{@{}lllll@{}}
\hline
 &  &  NO & IO & MO\\ 
\hline
\multirow{3}*{2nd octant vs.~1st} & $\log \mcB$ & $0.3$ &  $1.2$ & $0.7$ \\
  & $\Delta \AIC/2 $ & $-0.5$ &  $1$ & $0.5$ \\
  & $\Delta \chi^2 $ & $-0.9$ &  $2.0$ & $1.0$ \\ \hline
\multirow{3}*{Non-maximal vs.~maximal} & $\log \mcB$  & $-1.5$ &  $-1.2$ & $-1.3$ \\
  & $\Delta \AIC/2 $ & $-0.5$ &  $0.0$ & $0.0$ \\
 & $\Delta\chi^2 $ & $0.9$ &  $2.0$ & $2.0$ \\
   \hline
\end{tabular}
\end{center}
\caption{\it Model comparison for different assumptions on
  $\snq{23}$. Logarithms of Bayes factors,
  the comparable differences in the AIC, and differences in $\chi^2$ minima. 
The sign is chosen such that positive values correspond to preference for first mentioned assumptions in each case, \ie, the 2nd octant and non-maximal mixing, respectively. }
\label{tab:MCs23}
\end{table}
The second octant is weakly preferred over the first for the inverted
ordering, but not in the normal and the mixed orderings. 
Using the AIC, with the values also given in \tabref{tab:MCs23},
yields the same conclusions, although we remind the reader that
interpreting the AIC as a model likelihood should be done with great
care.  Due to the relatively bad predictivity of the assumption 
of non-maximal mixing, maximal
mixing is weakly preferred over non-maximal in all orderings. Note that $\Delta
\AIC/2$ can never be smaller than $-1$ in this case, and these numbers
close to that limit are simply saying that for no ordering is there any
preference for non-maximal mixing. 

If in the future the uncertainty on $\snq{23}$ keeps on being reduced while maximal mixing continues to be allowed,
at some point reducing the uncertainty further is pointless for the purpose of 
determining whether maximal-mixing is the correct model. 
Bayesian model comparison gives a
quantification of at which point this is the case, which is when the
evidence in favour of non-maximal mixing becomes strong.

\section{Exploring $\dcp$ and CP-violation}
\label{sec:deltacp}
\begin{figure}[t!]
\includegraphics[width=0.4\textwidth]{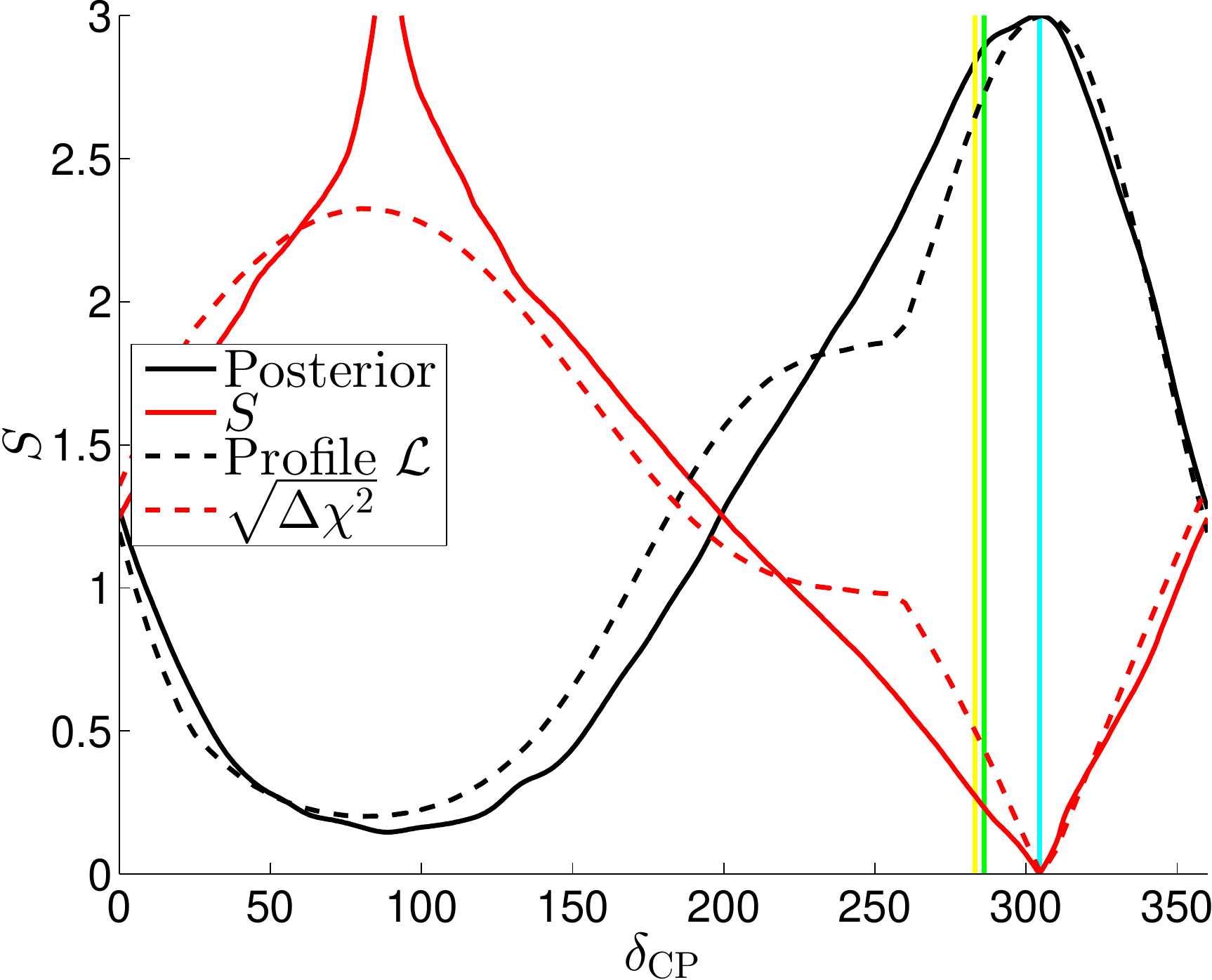}
\includegraphics[width=0.4\textwidth]{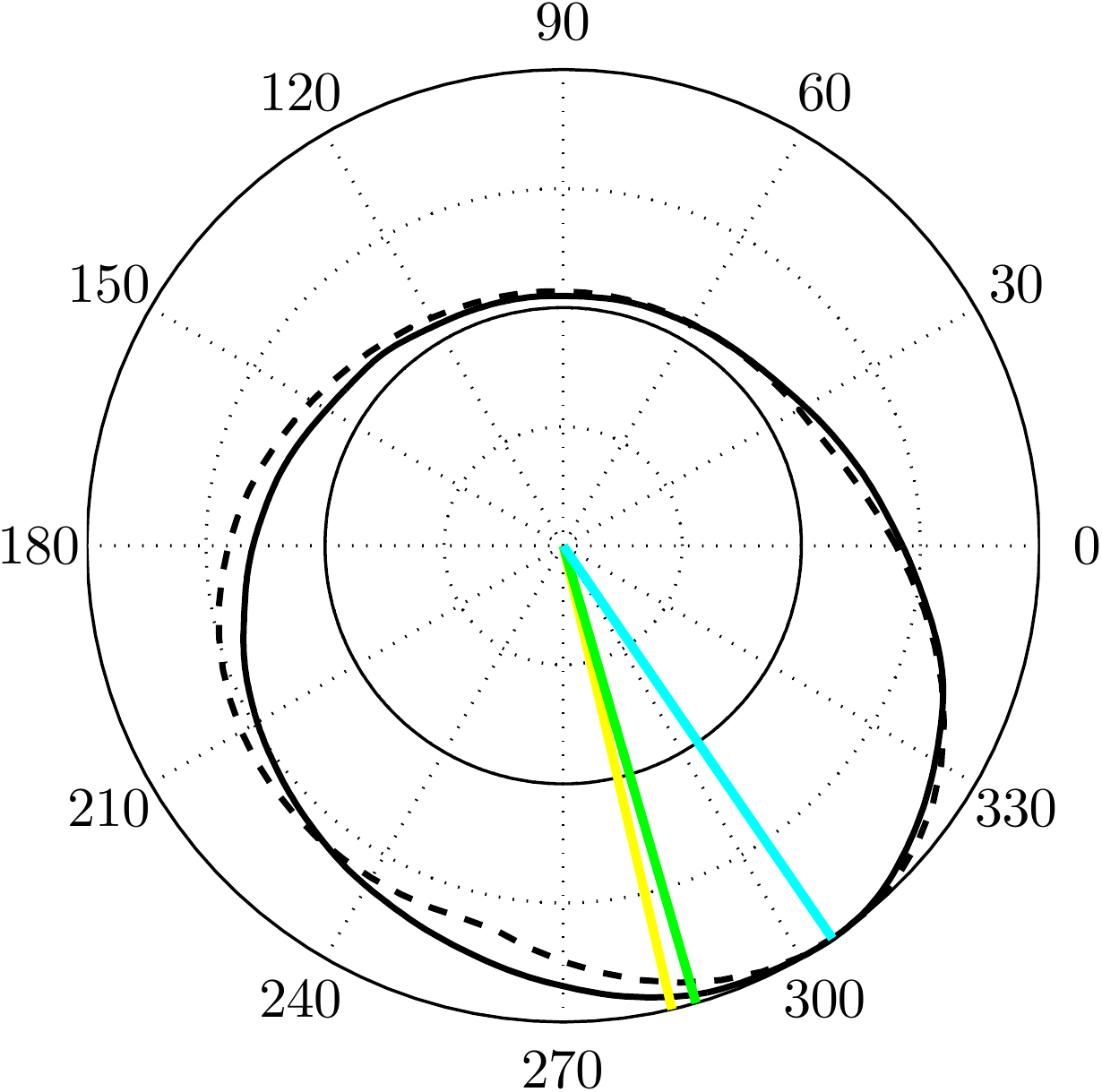}
\includegraphics[width=0.4\textwidth]{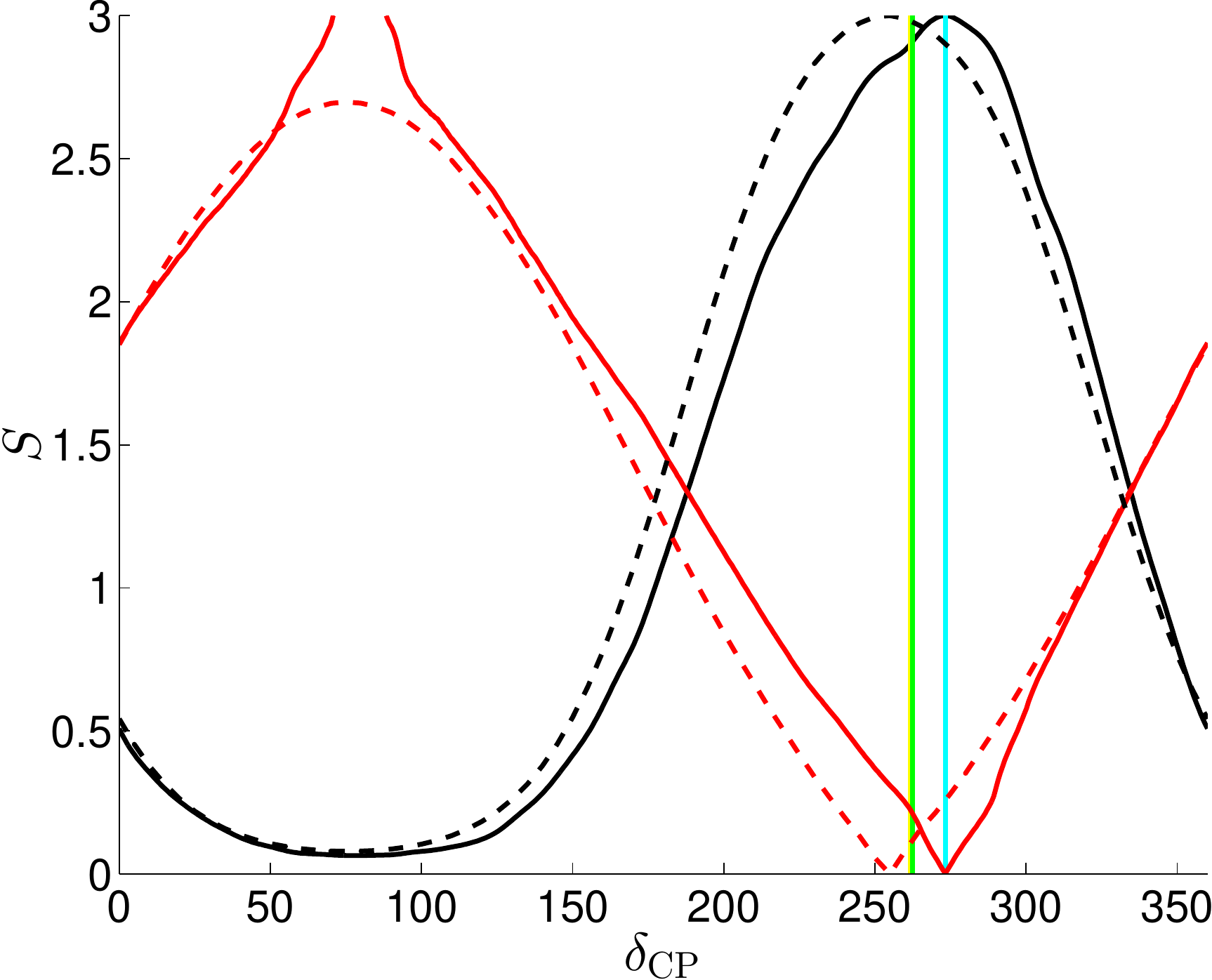}
\includegraphics[width=0.4\textwidth]{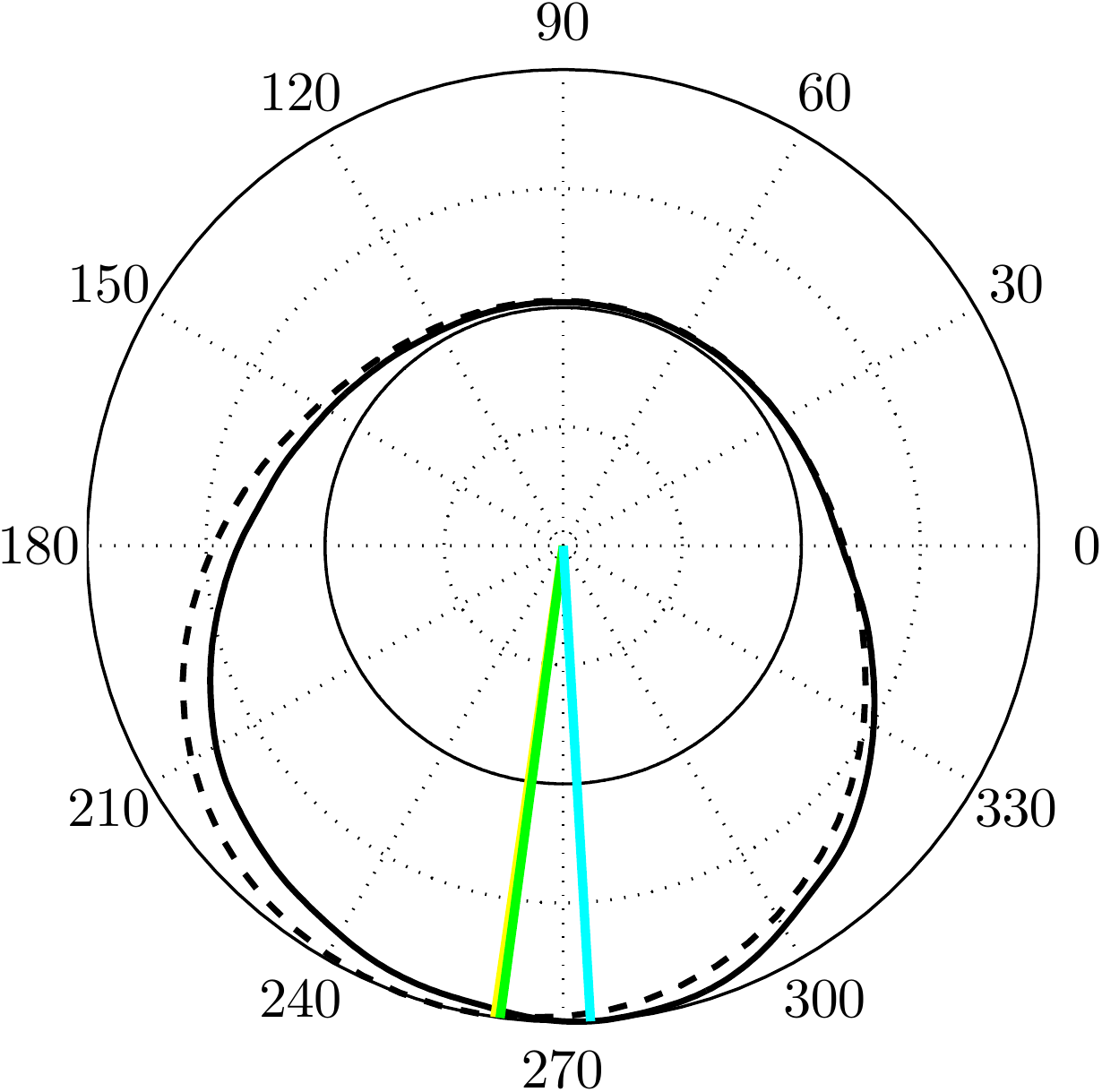}
\includegraphics[width=0.4\textwidth]{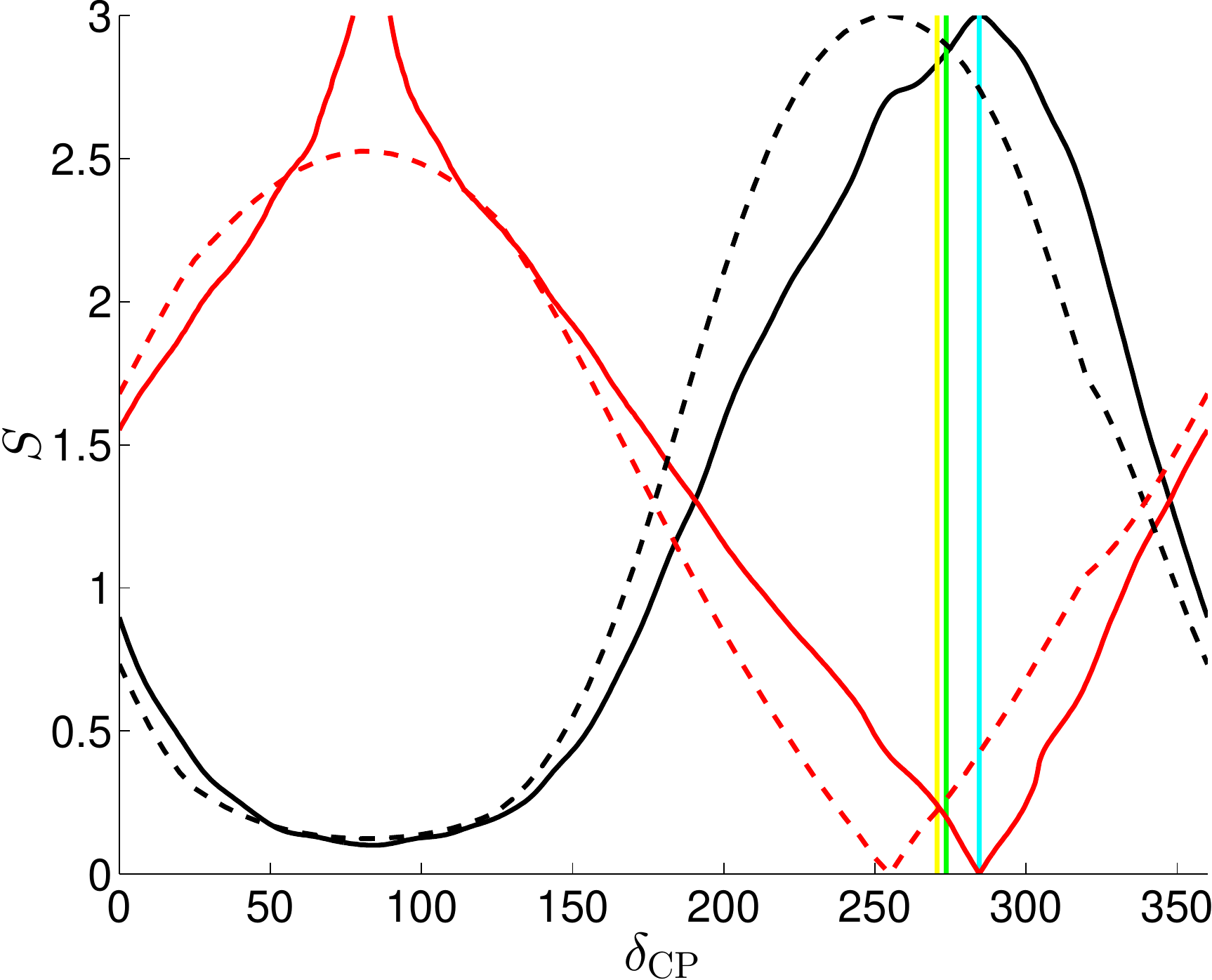}\hspace*{3cm}
\includegraphics[width=0.4\textwidth]{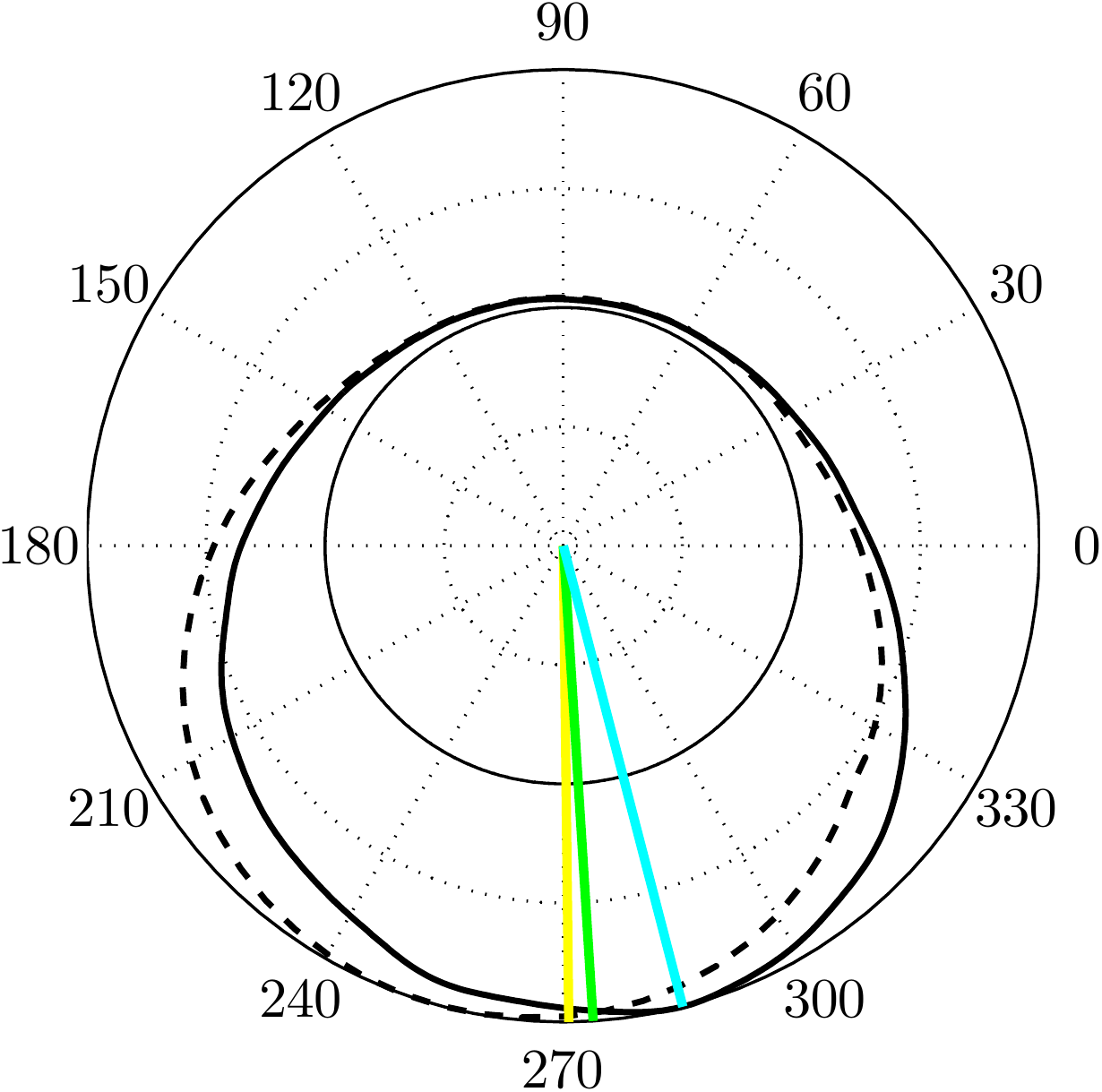}
\caption{Left plots: same as \figref{fig:snq23_comb} for $\dcp$.
  Right plots: Same as left plots, but with only posterior and
  profile likelihood and plotted in polar coordinates. For clarity,
  half of the maximal radius corresponds to zero function
  value. }\label{fig:dlt_comb}
\end{figure}

In this section we study the determination of $\dcp$ in more detail.
In the left panels of \figref{fig:dlt_comb} 
we plot the Bayesian marginal posterior distribution  of
$\dcp$ for all orderings
together with the $S$ of the credible intervals, 
as well as the profile likelihood
and $\sqrt{\Delta \chi^2}$. 
For NO, the marginal and profile
likelihoods have their maximum at about the same value of $\dcp$, but
for IO and MO, the Bayesian analysis prefers larger $\dcp$. Comparing
$S$ with $\sqrt{\Delta \chi^2}$, the difference is not that large,
apart from the shift just mentioned, and the fact that $S$ diverges
near $\dcp \simeq 90^\circ$, while $\sqrt{\Delta \chi^2}$ is bounded
by about $2.5$. 

In the right panels of \figref{fig:dlt_comb} the
marginal and profile likelihoods are plotted again, 
but in a polar coordinate system
which better reflects its circular nature.
We note that in a frequentist analysis the fact that $\dcp$ is a phase and a
circular, periodic variable will affect distributions of test
statistics \cite{Blennow:2014sja,Elevant:2015ska}. 
For the present data $\sqrt{\Delta \chi^2}$ is expected to be a poor
approximation of the frequentist significance, and typically the true
significance will be higher than the naive expectation. Hence,
\figref{fig:dlt_comb} does not give a direct comparison of frequentist
and Bayesian results.

In the Bayesian analysis, however, the circular nature of $\dcp$ does
not affect the posterior distributions or its interpretation.
Nevertheless, it still needs to be taken into account if one wants to
make summaries of the posterior in terms of point estimates such as
the mean, median, or measures of dispersion such as the standard
deviation. This is because the normal, linear definitions of these
quantities will depend on the arbitrary choice of origin for
$\dcp$~\cite{Jammalamadaka:2001,Mardia:2009,Fisher:1995}.

In this respect a useful summary of the distribution 
of $\delta_{\rm CP}$ is given by the first moment,
\be 
m_1 = \mean{e^{i \dcp}}, 
\ee
with $\mean{\cdot}$ denoting the mean 
(indeed, it is $e^{i \dcp}$ which enters the mixing matrix). 
The appropriate analogues of
the mean and median of $\dcp$ are the \emph{circular mean}  
and \emph{circular median}. The first one is given by the argument 
of the first moment,
\be 
\overline{\dcp} = \arg m_1 = \arg \mean{e^{i \dcp}},
\ee
while the second is defined as the endpoint closer to
mean of the diameter of the circle that has $0.5$ probability on each
of its sides.  These point estimates are summarized in 
\tabref{tab:dlt_pointest} together with the likelihood maxima, and
their values are plotted in \figref{fig:dlt_comb}. 

In what respects characterization of the dispersion, besides
the credible intervals, if one wants to have a  characterization
similar to that provided by the linear standard deviation, one can 
make use of  the fact that $R=|m_1|$ gives a reasonable measure of dispersion, 
with $R=0$ for a uniform distribution and $R=1$ for a degenerate one.  
However, it could be preferable and 
more easily interpretable to
have such a measure which is an expected deviation \emph{in
  radians}. Noting that the standard linear variance is the
expectation of the Euclidean distance squared from the mean, in
general one could use
\be 
V = \mean{d^2(\dcp,\overline{\dcp})}  
\ee
to obtain a dispersion, where $d$ is some metric on the circle. 
The usual linear metric $d(\alpha,\beta) = |\alpha - \beta|$ is not
invariant with respect to choice of origin, but one can  take instead
$d$ as the minimum arc length between $\alpha$ and $\beta$, also
called the great-circle distance. Hence, one can simply take $\sigma =
\sqrt{\mean{d^2(\dcp,\overline{\dcp})}} $ as the variance. 

Another metric one can use is the one inherited from the Euclidean embedding, 
\be 
d'(\alpha,\beta)^2=| e^{\im \alpha} - e^{\im \beta} |^2  =  
(\sin \alpha - \sin \beta)^2 + (\cos \alpha - \cos \beta)^2 = 
2(1-\cos(\alpha-\beta)). 
\ee
Then, the variance becomes
\be 
V = \mean{d'(\dcp,\oldcp)^2} = \mean{2 (1 - \cos(\dcp-\oldcp))} = 2(1-R), 
\ee
since $R = \mean{\cos(\dcp-\oldcp)}$. To get the equivalent deviation
as an angle away from the mean, we solve $V=2(1-\cos \sigma')$, giving simply
\be 
\sigma' = \arccos R,
\ee
which is then the deviation from the mean which has the same distance
squared as the expectation over the distribution.  These measures of
dispersion, together with the corresponding credible intervals, are show in
\tabref{tab:dlt_unc}.

\begin{table}[h]
\begin{center}
\begin{tabular}{@{}lllll@{}}
\hline
	Ordering	& Global max $[{}^\circ]$	   & Max of  $\lhood_{\rm marg}$ $[{}^\circ]$ 	& mean  $[{}^\circ]$& median $[{}^\circ]$\\ 
\hline
 NO	  			&	306							& $304$										&  $289$ 			& $286$  \\ 
 IO	  			&	254							& $273$ 									& $262$  			& $262$ \\ 
 MO	  			&	254							& $289$  									& $271$  			& $272$ \\ 
 \hline
\end{tabular}
\end{center}
\caption{\it Point estimates of $\dcp$. The mean and median are the corresponding circular quantities. \label{tab:dlt_pointest}}
\end{table}

\begin{table}[h]
\begin{center}
\begin{tabular}{@{}llllll@{}}
\hline
		Ordering	& Method	&$\sigma/\sigma' [{}^\circ]$& $1 \sigma$ CI $[{}^\circ]$	& $ 2 \sigma$ CI $[{}^\circ]$& $ 3 \sigma$ CI $[{}^\circ]$\\ 
\hline
 \multirow{2}*{NO} 	& Bayes		& $65/58$					& $ [223,350]$  				&  $ [42,139]^c$ 	& $[84,94]^c$   \\ 
  					&$\chi^2$ 	& -							& $ [234,346]$  				&  $ [33,131]^c$ 	&  - \\
 \multirow{2}*{IO}	& Bayes		& $56/51$					& $[207,319]$   				&  $ [9,146]^c$ 	& $[70,90]^c$ \\ 
					& $\chi^2$	& -							& $[192,317]$   				&  $ [8,142]^c$ 	& - \\ 
 \multirow{2}*{MO}	&Bayes		& $61/55$ 					& $[211,333]$					& $ [28,144]^c$		& $[76,90]^c$  \\
				 	& $\chi^2$	& -							& $[192,317]$					& $ [16,142]^c$		& -  \\
  \hline
\end{tabular}
\end{center}
\caption{\it Measures of dispersion and credible intervals. Here, $I^c$ is the complement of $I$, \ie, all values of $\dcp$ not contained in $I$. \label{tab:dlt_unc}}
\end{table}

\begin{figure}[t!]
\includegraphics[width=0.4\textwidth]{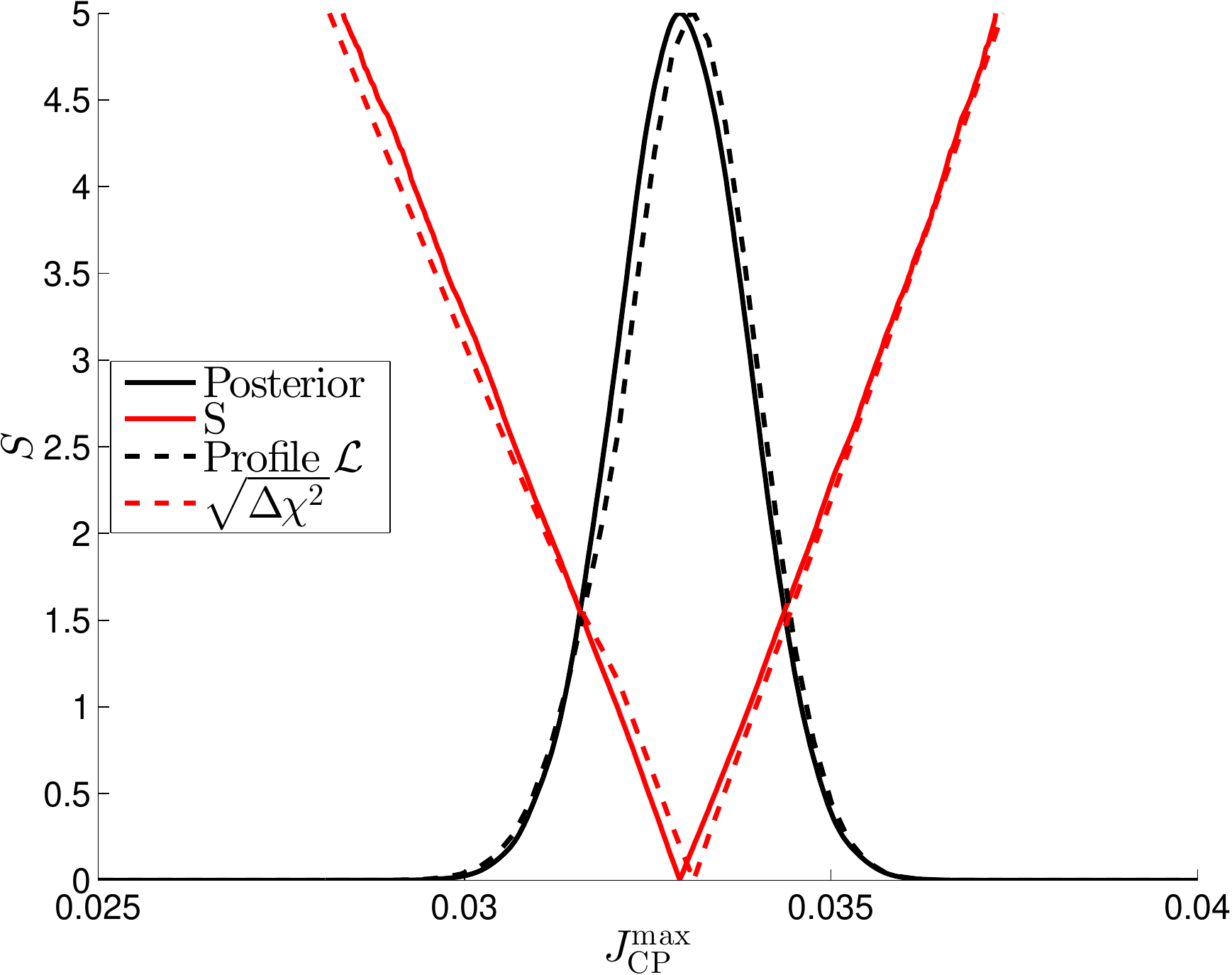}
\includegraphics[width=0.4\textwidth]{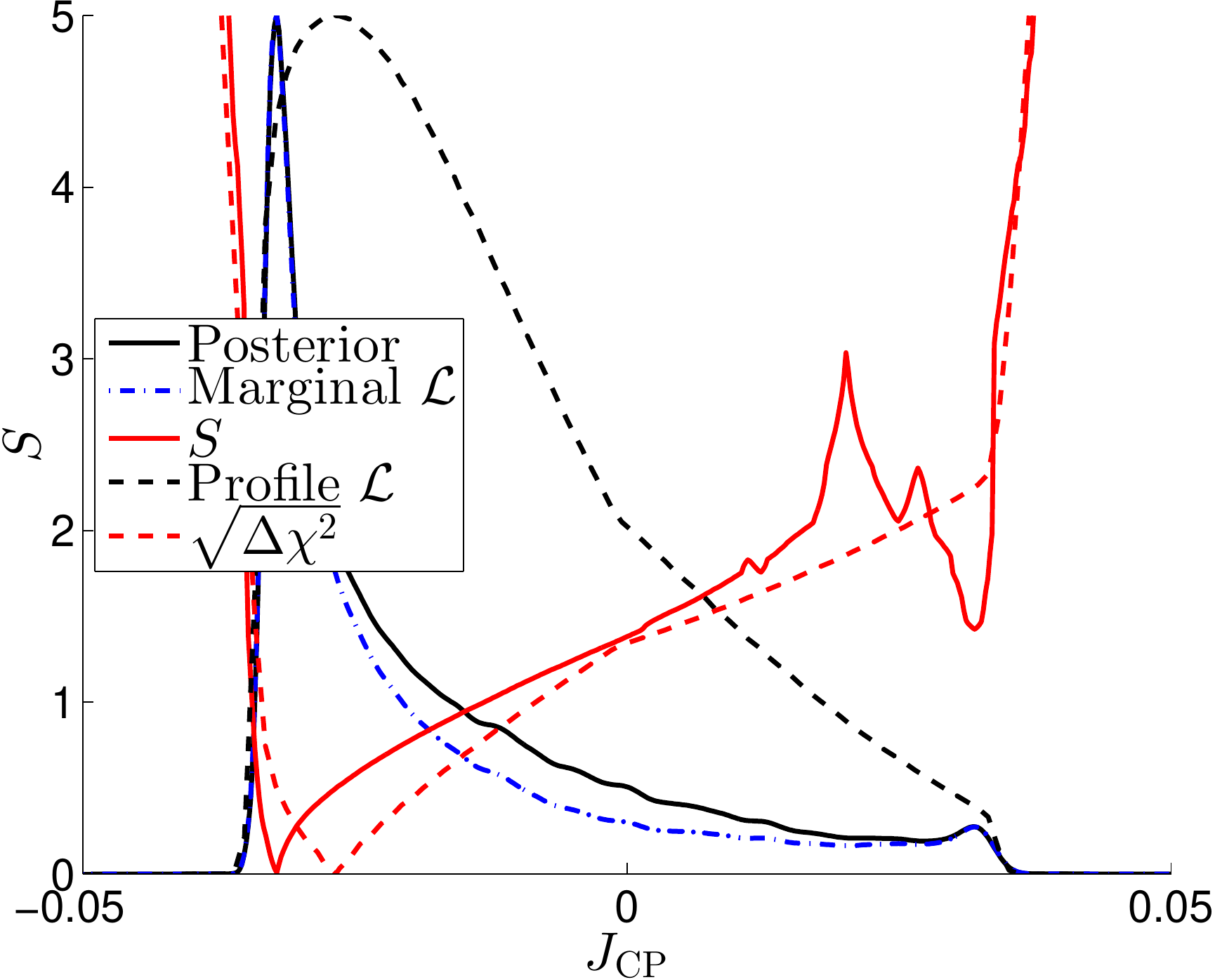}
\includegraphics[width=0.4\textwidth]{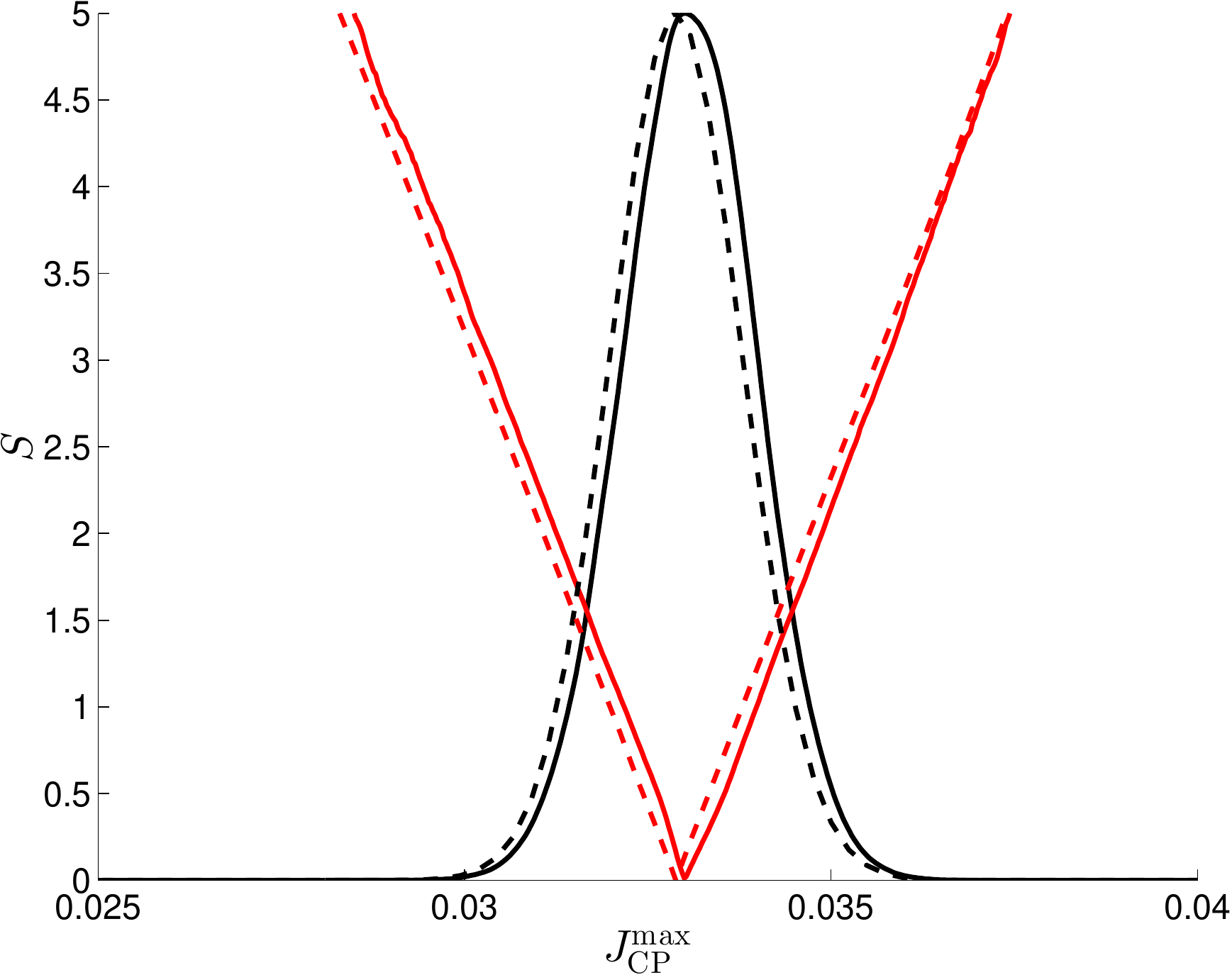}
\includegraphics[width=0.4\textwidth]{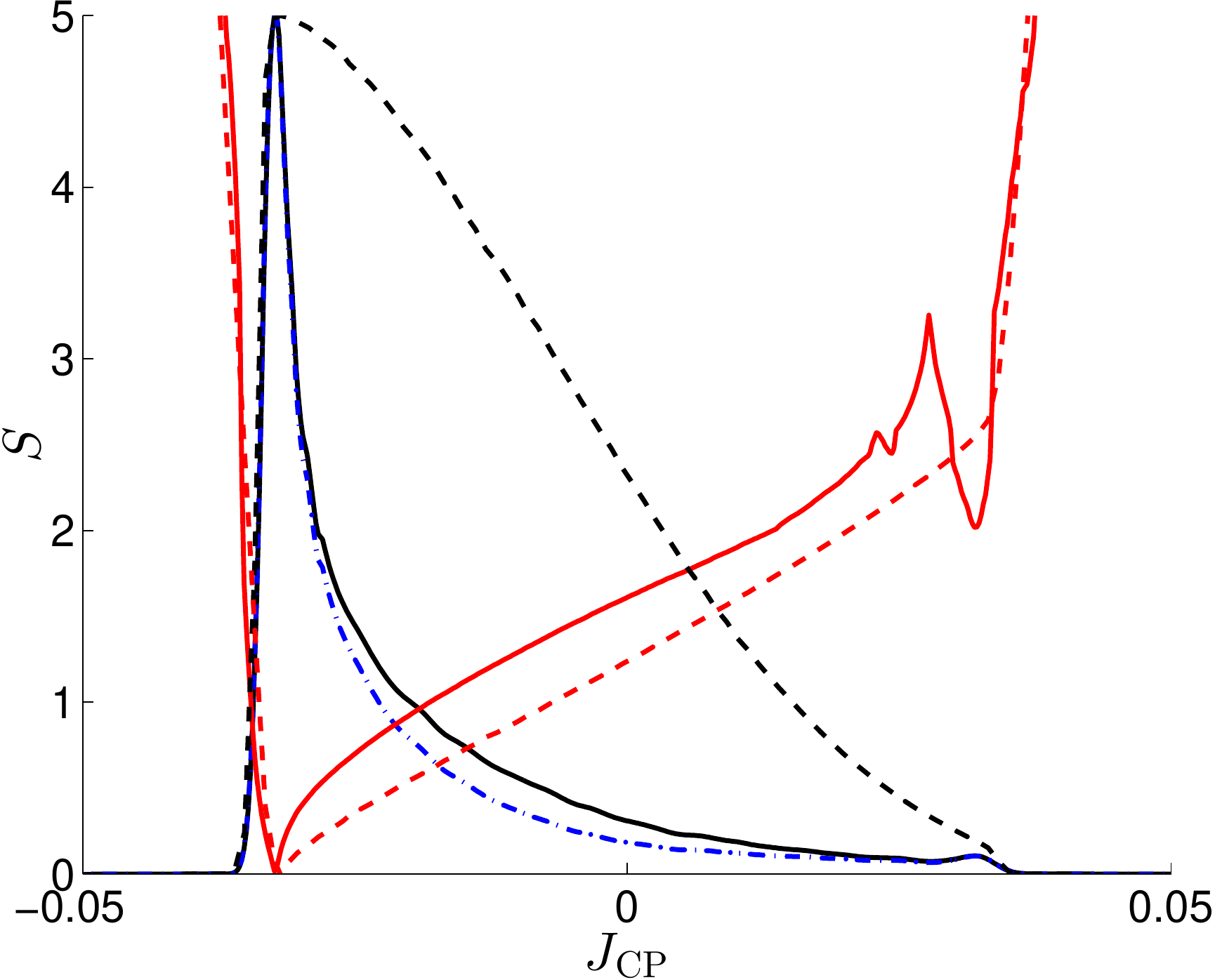}
\includegraphics[width=0.4\textwidth]{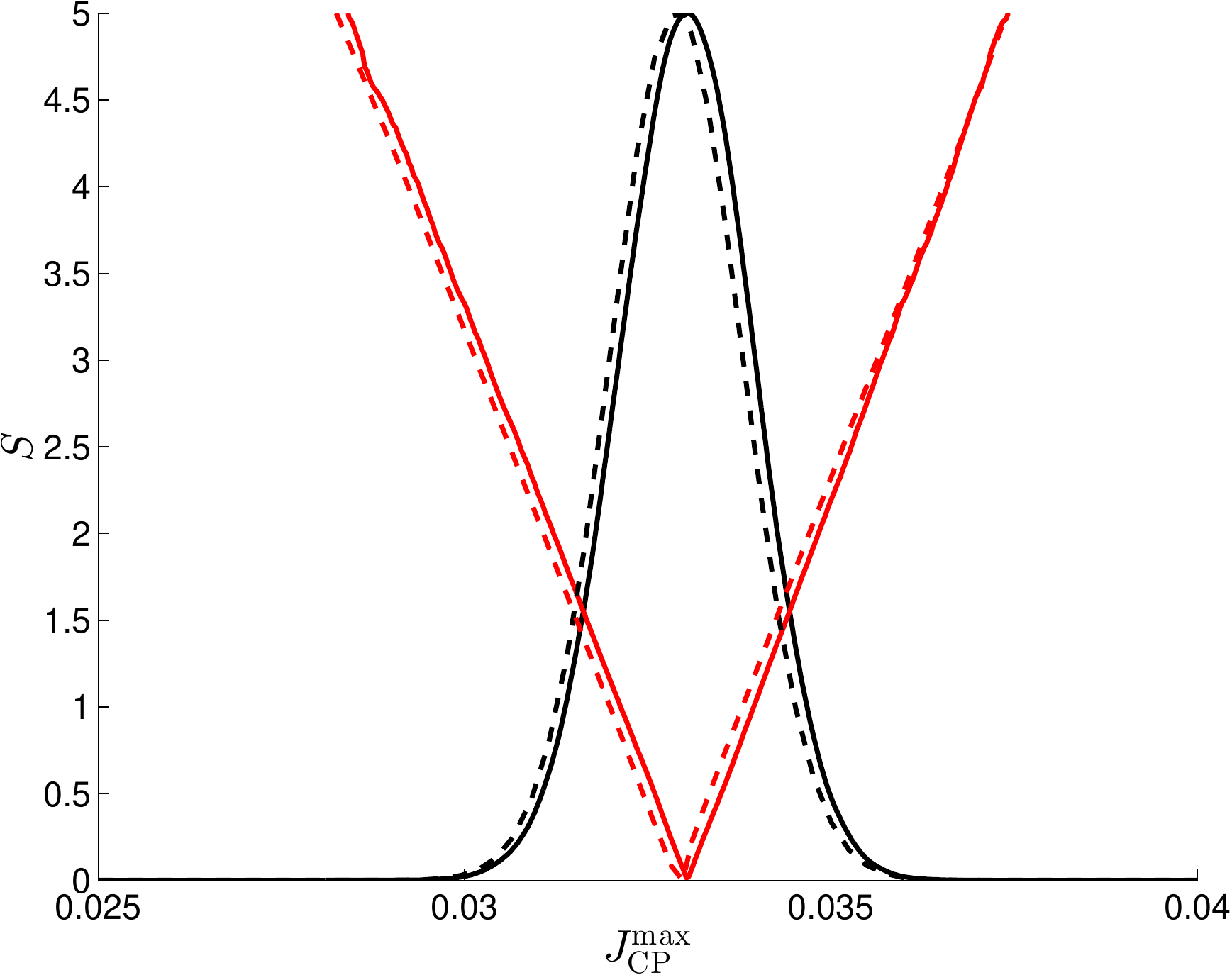}\hspace*{3cm}
\includegraphics[width=0.4\textwidth]{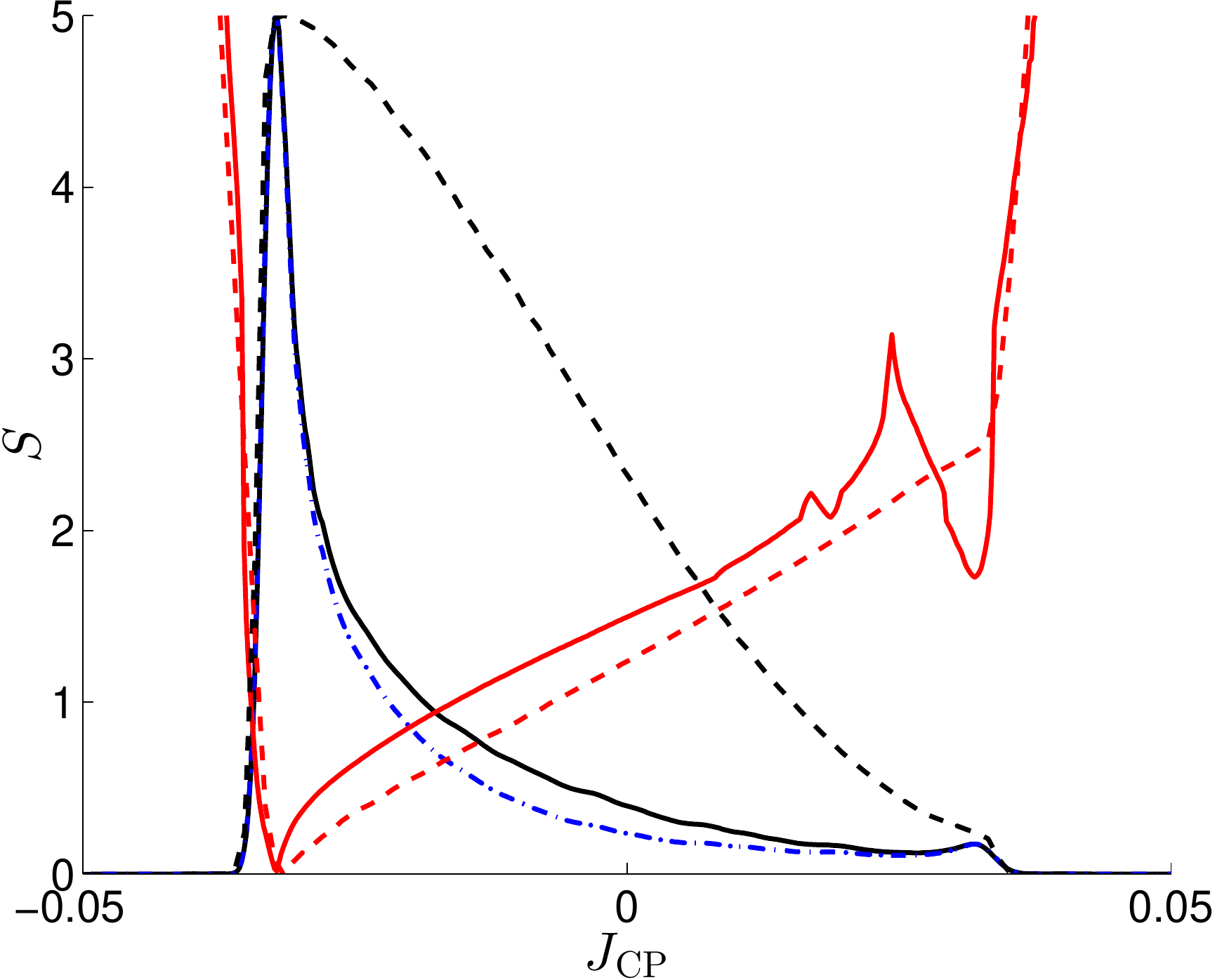}
\caption{Jarlskog invariant and its maximal value for all
  orderings. NO (top), IO (middle), MO (bottom).}\label{fig:JCP}
\end{figure}

The presence of CP violation can also be 
studied in terms of the Jarlskog invariant, $J_{\rm CP}$,  which, 
in the standard parameterization, is given by
\be 
J_{\rm CP} = J^{\rm max}_{\rm CP} 
\sin \dcp = c_{12} s_{12} c_{23} s_{23} c_{13}^2 s_{13}  \sin \dcp.
\ee 
We plot in \figref{fig:JCP} the Bayesian marginal posterior distribution  
of  $J_{\rm CP}$ and $J^{\rm max}_{\rm CP}$  for all orderings
together with the $S$ of the credible intervals, as well as the profile 
likelihood and $\sqrt{\Delta \chi^2}$.  We note that 
these are derived parameters, and so their priors and posteriors are
determined by those of the free oscillation parameters. In particular,
their priors are not exactly uniform. For  $J^{\rm max}_{\rm CP} $ 
(the left panels) the prior is very close to uniform, 
and from the figure we see that it is so well constrained that it
is perfectly Gaussian and agrees with the profile
likelihood. 

For $J_{\rm CP} $  (right panels of \figref{fig:JCP}), we plot both
the posterior and the marginal likelihood, and we observe a difference,
although it is not very large. A much larger difference is observed 
between these and the profile likelihood, which translates into a 
difference in the corresponding CL's ($S$ and $\sqrt{\Delta\chi^2}$). 
However, this difference is much smaller 
than one could naively expect form the differences in
posterior versus the profile likelihood, 
the reason  for this being that the Bayesian results are a function 
of the total probability contained in a region, and the sharp peak in the 
posterior still contains relatively little probability.

That the posterior of $J_{\rm CP} $ shows peaks towards the edges of
the distribution is simply because the density of $|\sin \dcp|$ is
larger for those values.  This is  not canceled out in the marginal
likelihood because $J^{\rm max}_{\rm CP} $ has a broad prior, which
means that so has $J_{\rm CP}$. Of course, the symmetry around $J_{\rm
CP} = 0$ is broken by the information on $\dcp$ supplied by the data,
which then means 
that negative values of $J_{\rm CP}$ are preferred,
and more strongly so than in the $\chi^2$ analysis.  Note that since
we do not have any freedom left in choosing our priors on the
oscillation angles and phase, this is in some sense a robust
consequence of using consistent Bayesian inference.

\subsection{CP-violation vs CP-conservation}
In the same way as maximal mixing, one can consider either exact
CP-conservation as a possible scenario, or alternatively simply
CP-conservation as a very good approximation, and compare the
models:
\begin{itemize}
\item $M_{\rm CPC}^1$: $\delta=0$
\item $M_{\rm CPC}^2$: $\delta=180^{\circ}$
\item $M_{\rm CPC}: M_{\rm CPC}^1$ or $M_{\rm CPC}^2$ (with equal priors)
\item $M_{\rm CPV}$: $\delta  \in [0^{\circ}, 360^{\circ}] 
\setminus \{0^{\circ}, 180^{\circ}\} $, with prior $\pi(\dcp) = 1/360^{\circ}$.
\end{itemize}

Note that these assumptions on CPC and CPV are unambiguously defined
in the sense that they do not depend on a parameterization, and that
the prior on $\dcp$ in $M_{\rm CPV}$ is uniquely given by the Haar
measure. Hence, there is essentially no flexibility remaining in the
choice of prior.  Due to this fact and the compact nature of the
parameter space, the normal pitfalls of model comparison, \ie, the
potentially large and prior dependent penalty acquired for additional
parametric complexity, are avoided, or at least heavily mitigated.

This unusually robust (fixed in size) and small penalty for the
additional parameter means that the Bayesian analysis is expected to
be more powerful at detecting CPV than it normally is at detecting a
new physical effect. Hence, when comparing with a $\chi^2$ analysis, a
smaller significance or value of $\Delta \chi^2 $ than normally should
be needed for robust, Bayesian, detection of CPV. Equivalently, a
certain value of $\Delta \chi^2$ would lead to a stronger Bayesian
evidence of CPV than what the same $\Delta \chi^2$ would yield in a
different setting. 

Interestingly, also the true frequentist significance of CP-violation
is expected to be stronger than the naive expectation
\cite{Elevant:2015ska}, although the details depend significantly on
the (unknown) value of $s^2_{23}$ assumed to be true \footnote{
This is the case  particularly
in analysis of the current data, where sensitivity to 
$\dcp$ is poor.  However for more sensitive data the behaviour is expected
to become more Gaussian \cite{Blennow:2013oma}.}.
This does not
happen in a Bayesian analysis, which also does not depend on any
distributions of test statistics under repeated experiments, but only
on likelihood of the data which was actually observed.

The likelihoods of the different assumptions on $\dcp$, in the usual
form of logarithms of Bayes factors, $\log (\ev/\ev_{\rm CPV})$
relative to $M_{\rm CPV}$ are shown in \tabref{tab:MCCPV}, together
with the AIC and difference in $\chi^2$. 
Although technically CP-violation is preferred
in all cases, in none of the cases is the evidence even weak.
Notice also that since $\dcp$ is relatively
unconstrained, the preference for CPV is even smaller using the AIC
than in the Bayesian analysis.

\begin{table}[h]
\begin{center}
\begin{tabular}{@{}lllll@{}}
\hline
 &  									& NO 		& IO 		& MO\\ 
\hline
\multirow{3}*{$M_{\rm CPC}^1$}& $\log \mcB$ 		& $-0.1$	
&  $-0.8$	& $-0.4$ \\
					&  $\Delta \AIC/2$	&  $0.1$
&  $-0.7$ 	& $-0.4$ \\
					&  $\Delta \chi^2$	&  $-1.8$
&  $-3.4$ 	& $-2.8$ \\	 \hline

\multirow{3}*{$M^2_{\rm CPC}$}& $\log \mcB$ 		&  $-0.4$ 	
&  $-0.1$ 	& $-0.2$ \\
					&  $\Delta \AIC/2$	&  $0.1$
&  $0.3$ 	& $0.3$ \\
					&  $\Delta \chi^2$	&  $-1.8$ 
&  $-1.5$ 	& $-1.5$ \\ \hline

\multirow{3}*{$M_{\rm CPC}$}&$\log \mcB$& $-0.2$ 	& $-0.4$  	
& $-0.3$  \\
					&  $\Delta \AIC/2$	&  $0.1$
&  $0.3$ 	& $0.3$ \\
					&  $\Delta \chi^2$	&  $-1.8$ 
&  $-1.5$ 	& $-1.5$ \\
\hline 
\end{tabular}
\end{center}
\caption{\it Model comparison for different assumptions on
  $\dcp$. Logarithms of Bayes factors relative to $M_{\rm CPV}$, the
  comparable differences in the AIC, and differences in $\chi^2$ 
  as $\Delta\chi^2 = \chi^2_{\rm M_{CPV}} - \chi^2_{\rm M^i_{CPC}}$.  
For all variables,  positive values would indicate preference 
of the corresponding $M^i_{\rm CPC}$ over  $M_{\rm CPV}$.  }
\label{tab:MCCPV}
\end{table}

\section{Correlation between $s_{23}^2$ and $\dcp$}
\label{sec:corre}
\begin{figure}[h]
\begin{center}
\includegraphics[width=0.4\textwidth]{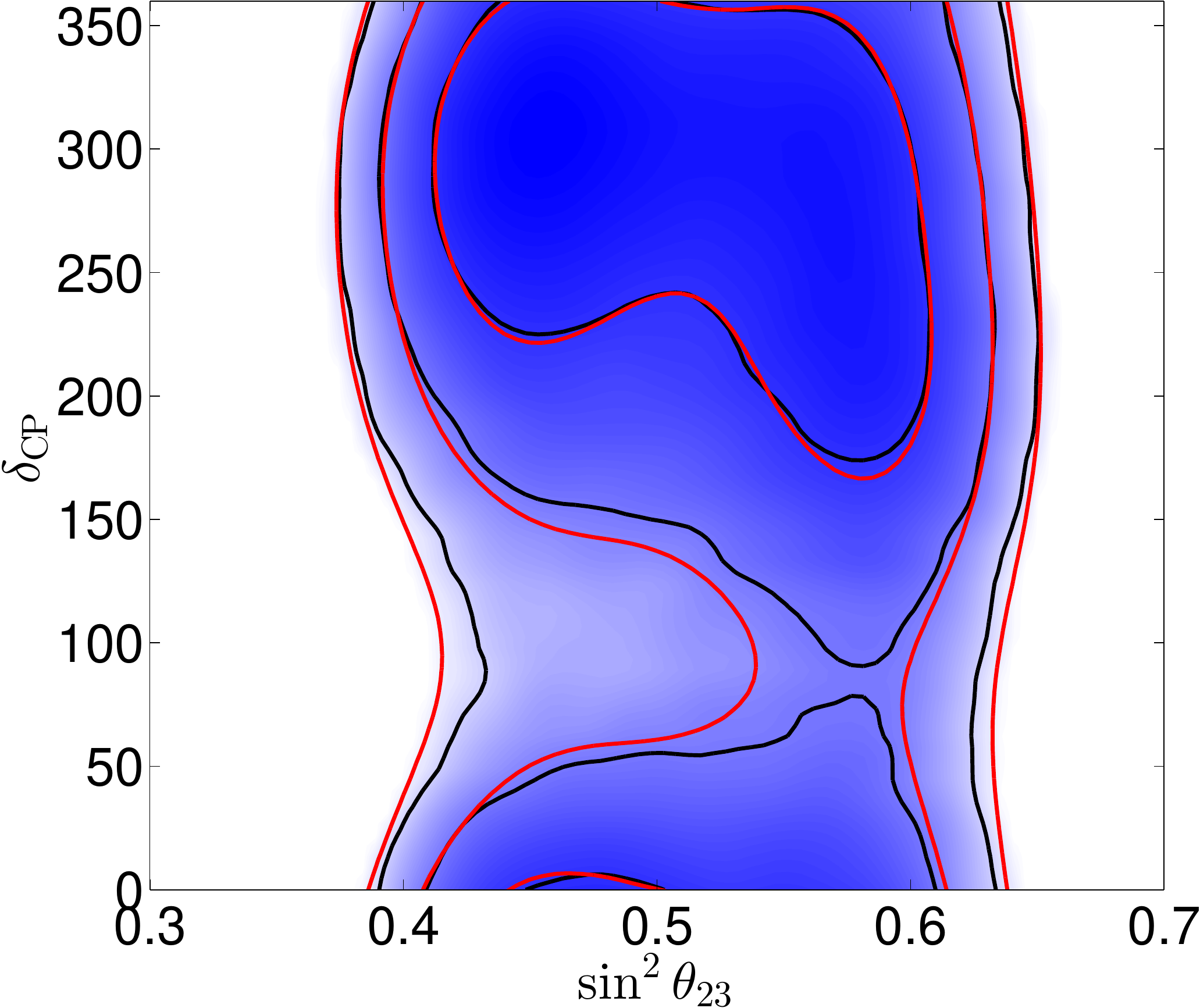}
\includegraphics[width=0.4\textwidth]{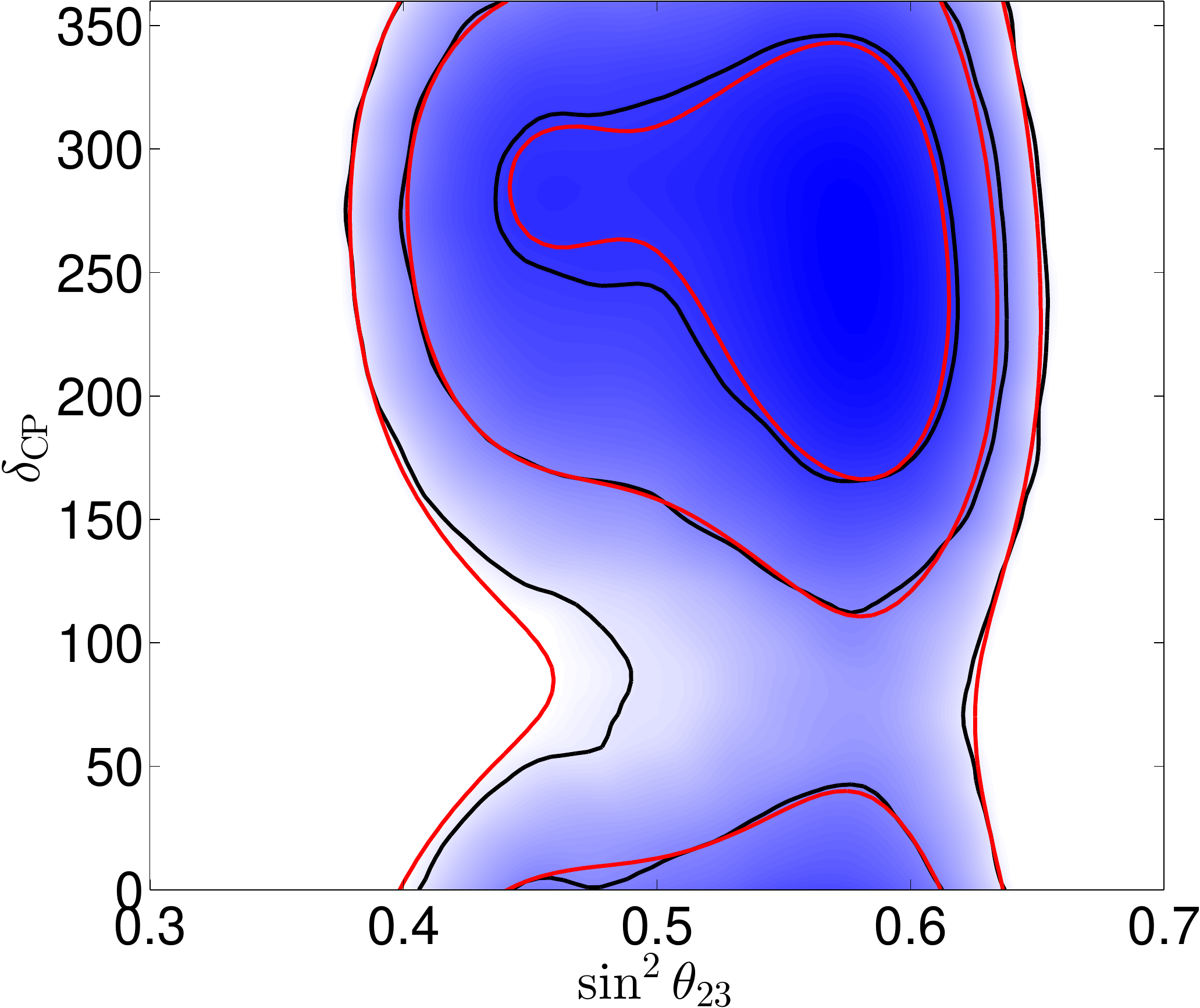}
\includegraphics[width=0.4\textwidth]{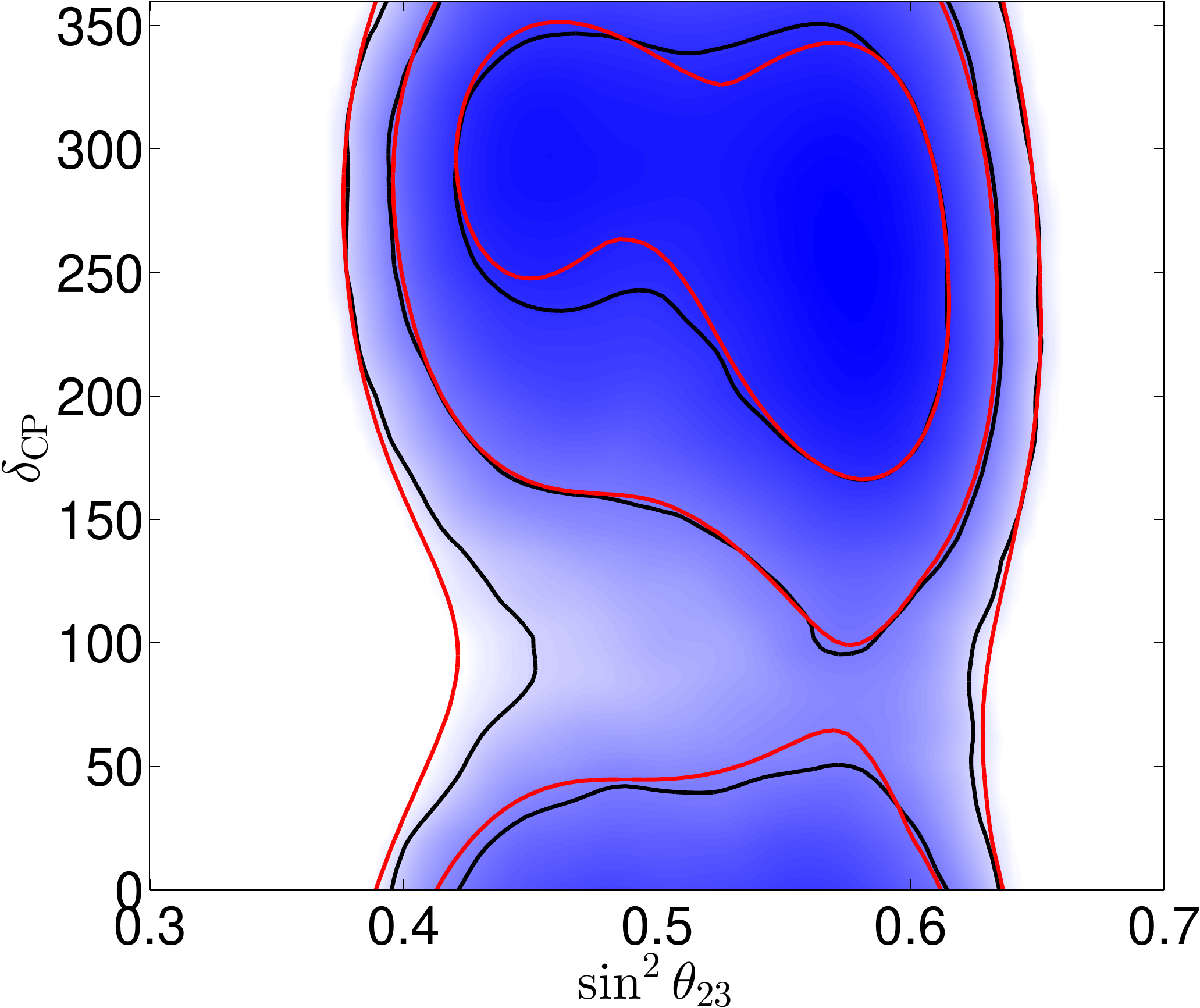}
\end{center}
\caption{Posterior in the $\snq{23} - \dcp$ plane (blue shading), $1
  \sigma$, $2 \sigma$, $3 \sigma$ credible regions (black) $\chi^2$
  contours (red dashed). NO (top left), IO (top right), MO (bottom).  }
\label{fig:2D}
\end{figure}

In this section we discuss the possible quantification of the 
correlation between $\sin^2\theta_{23}$ and $\dcp$. 
The posterior in the $\snq{23} - \dcp$ plane for all the orderings is
plotted in \figref{fig:2D}, together with the credible regions and
$\chi^2$ contours. Although the difference between the Bayesian and
$\chi^2$ analysis does not appear to be extremely large, there are
some things which a Bayesian analysis makes possible which cannot be
done in a $\chi^2$ analysis. In particular, as seen in the figure, 
it is clear that $\snq{23}$ and $\dcp$ are
not independent, and it will be interesting to quantify if the
degeneracy between them is something which persists in future
experiments.  In a $\chi^2$ analysis, quantifying the \qu{correlation}
between two parameters is typically limited to fitting a two
dimensional Gaussian at the best-fit point.  In a Bayesian analysis,
global measures of association such as the standard Pearson
product-moment correlation coefficient are available.  However, this
one only measures linear association, and is hence less useful when
there are non-linear trends involved, including multi-modality. In
particular it is possible for two highly dependent variables to have
very small value of the Pearson correlation.  Furthermore, in the
present case, it fails in an even worse manner since the Pearson
correlation is not circular invariant, i.e., its value depends on the
arbitrary choice of origin for $\dcp$. 
In what respects $\theta_{23}$
one can treat $\theta_{23}$ as circular variable or use instead the 
linear variable $\snq{23}$.  

So let us focus on how to define
a correlation coefficient which can overcome these limitations. 
Typically a correlation coefficient will aim to quantify how much of
the variation in one variable can be explained by the variation in
another one.  For example, to what extent the linear relation
$\mean{Y|X=x} = ax+b$ is responsible for the variation in $Y$ (which leads
to the standard Pearson correlation coefficient).  Similarly
one can consider circular-circular associations between two circular
variables $\Theta$ and $\Phi$ (in this case 
$\dcp$ and $\theta_{23}$),  circular-linear association, 
predicting the expectation of $\Theta$, given $X = x$ , 
or linear-circular association, predicting
the expectation of $X$, given $\Theta=\theta$
(in these cases  $X=\sin^2\theta_{23}$).

Many measures of correlation involving circular variables already exist
in the literature
(see \cite{Jammalamadaka:2001,Mardia:2009,Fisher:1995,Jones2006176}).
For two circular variables a  simple one is 
\be \rho_{\rm cc} =
\frac{ \mean{ \sin(\Theta - \bar{\Theta})\sin(\Phi -
    \bar{\Phi})}}{\sqrt{\mean{ \sin^2(\Theta - \bar{\Theta}) } \mean{
      \sin^2(\Phi - \bar{\Phi})} }}, 
\label{eq:r_cc} 
\ee 
where the bar denotes the circular mean. This has many properties in 
common with the linear version, such as being confined to the interval 
$[-1,1]$, it is
zero if the variables are independent, and it numerically agrees with 
the linear version for concentrated distributions.

An alternative, but slightly more complex, correlation coefficient 
for two circular variables is
the T-linear one of \refcite{Fisher:1983}, 
\be \rho_{\rm T} = \frac{
  \mean{ \sin(\Theta_1 - \Theta_2)\sin(\Phi_1 - \Phi_2)}}{\sqrt{\mean{
      \sin^2(\Theta_1 - \Theta_2) } \mean{ \sin^2(\Phi_1 - \Phi_2)}
}}, 
\ee 
where $\Theta_1$ and $\Theta_2$ are treated as two independent
copies of $\Theta$, and similarly for $\Phi$.

Also for linear-circular association, one can split the circular variable into
its sine and cosine and consider the multiple correlation coefficient
between $X$ and $(\sin \Theta, \cos \Theta)$, giving 
\be 
\rho^2_{\rm
  lc} 
= \frac{ \rho^2_{xc} + \rho^2_{xs} - 2
  \rho_{xs}\rho_{xc}\rho_{cs} }{1- \rho^2_{cs}},
\ee with $\rho_{xc} =
\rho(x,\cos y)$, $\rho_{xs} = \rho(x,\sin y)$, $\rho_{cs} = \rho(\cos
y,\sin y)$ being standard linear coefficients. We notice that 
being defined by a square, only  $|\rho_{\rm lc}|$ is known 
and hence gives no information on the ``sign'' or ``direction'' 
of the association.

While the above measures of association overcome the problem of 
the circular invariance they are still only sensitive to a limited kind of association, and it is possible for them to be zero
even when the variables are highly dependent on each other. It could
hence be of interest to have a measure which can quantify any type of
dependence, and which will only be zero when the variables are
independent.  Such a measure, based on information theory, is the
\emph{mutual information}
\cite{Murphy:2012,PhysRevE.69.066138,Moddemeijer1989233,Steuer01102002}. This
is information gained by knowing the full distribution $P(x,y)$ rather
than only the marginal distributions $P(x)$, and $P(y)$, or
equivalently, the average information gained on $X$ by knowing the
value of $Y$ (and vice versa). This can be expressed as the so-called
Kullback-Liebler divergence between $P_{X,Y}$ and the product $P_X
P_Y$,
\be 
I(X,Y) = \int P(x,y) \log \frac{P(x,y)}{P(x) P(y)  }  \df x \df y. 
\ee
Using the natural logarithm gives the result in nats, while one
obtains the results in bits by using base 2.  It holds that $I(X,Y)
\geq 0$ with equality if and only if $X$ and $Y$ are
independent. Next, in order to make the connection with
the standard correlation coefficient, we note that for a
two-dimensional Gaussian distribution (for which no correlation is
equivalent to independence), $I=\log (1/\sqrt{1-\rho^2})$, and so we
define
\be 
\rho^2_{I} =1-e^{-2 I}. 
\ee
We now have constructed a correlation coefficient 
which is independent of any
boundary conditions on the variables and is invariant under arbitrary
univariate redefinitions of $x$ and $y$ (which the others are
not). As the previous coefficients it also reduces to the 
standard Pearson coefficient   in the limit of a concentrated Gaussian 
distribution.
However, like $|\rho_{\rm cl}|$, it only measures the degree of 
dependence, but not any \qu{direction} of the association.

Our estimates of the different correlation coefficients are given in
\tabref{tab:corr}.\footnote{We note that large biases in the
  estimation of the mutual information may occur
  \cite{PhysRevE.69.066138,Moddemeijer1989233}.  As before we use
  kernel density estimate of the densities, similar to
  \refcite{Steuer01102002}, and our very large sample sizes ensures an
  accurate estimate.}  
For all
measures, we find stronger correlation in NO than in IO, typically
significantly so (with the exception of $\rho_{\rm T}$). Furthermore,
the two signed circular-circular measures have significantly smaller
absolute values than the others, and for these we also find that in MO
the correlation is actually larger than both NO and IO, which is not
the case for the others. We note that all are smaller than or equal in
size of $|\rho_I|$. This is somehow expected as $|\rho_I|$, in some sense, 
measures \qu{all} the dependence between $\dcp$ and $\snq{23}$.

\begin{table}[h]
\begin{center}
\begin{tabular}{@{}llll@{}}
\hline
 						& NO  	& IO 	& MO \\ 
\hline 
$\rho_{\rm cc}$  			&$-0.20$&$-0.15$& $-0.21$   \\ 
$\rho_{\rm T}$  			& $-0.14$  &  $-0.13$ & $-0.16$     \\ 
$|\rho_{\rm cl}|$  			& $0.27 $	& $0.16$& $0.23$   \\ 
$|\rho_I|$  				&$0.30$	&$0.18$& $0.26$  \\ 
\hline 
\end{tabular}
\end{center}
\caption{\it Different correlation coefficients between $\snq{23}$ and $\dcp$.}
\label{tab:corr}
\end{table}

\section{Summary}
\label{sec:conclusions}
We have presented the results of a Bayesian global analysis of 
solar, atmospheric, reactor and accelerator neutrino data
in the framework of three-neutrino oscillations and compared
them with those from the standard $\chi^2$ analysis in NuFIT 2.0~\cite{nufit}. 
The results are summarized \figref{fig:param_NO} for
NO, \figref{fig:param_IO} for IO, and \figref{fig:param_MO} for MO
where we compare the relevant Bayesian quantities (the posterior
distribution and two-dimensional Bayesian credible regions) 
with the profile-likelihood and the two dimensional $\chi^2$ 
allowed regions.

We found that the four parameters $\Delta m^2_{3\ell}$, $\Delta
m^2_{21}$, $\snq{12}$, and $\snq{13}$, are well-measured and their
posterior distributions are Gaussian to a very good approximation.  The
corresponding Bayesian credibility intervals at a given CL are also
very similar to the $\chi^2$ allowed regions at the same CL, as seen
in Table~\ref{tab:paramdet}.

We found some differences between the results of the $\chi^2$ and Bayesian 
analysis where $\dcp$ or $\snq{23}$ are involved. In particular, the 
marginalization over $\dcp$ pulls the bulk of the posterior of $\snq{23}$ 
more into the second octant which has some effect on the ranges of 
parameter estimates and the 
quality of the description between octants. 
We study the determination of $\theta_{23}$ in more
detail in Sec.~\ref{sec:theta23} and we conclude that the Bayesian analysis 
generally prefer the second octant more so than the $\chi^2$ analysis, 
in particular for NO. The credible and confidence levels differ in the vicinity
of the two peaks but both peaks are within the $2 \sigma$ regions. 
Altogether the low-credibility Bayesian regions are 
larger than the small-$\chi^2$ regions, while the high-credibility Bayesian 
regions are smaller than the large-$\chi^2$ ones. 

In what respects the present determination of  $\dcp$, presented in 
Sec.~\ref{sec:deltacp}, we found that for NO, the marginal and profile
likelihoods have their maximum at about the same value of $\dcp$, but
for IO and MO, the Bayesian analysis prefers slightly larger values of 
$\dcp$. Also, unlike the $\chi^2$ interval, the 3$\sigma$ Bayesian credible 
interval do not contain the full range of $\dcp$ but some values near
$\pi/2$ are not included. We have also introduced and quantified two 
measures of the dispersion of $\dcp$ equivalent to the linear standard 
deviations but valid for a circular variable.  

In addition, we have studied the Jarlskog invariant, $J_{\rm CP}$, as well as its
maximal value over $\dcp$ and find that the posterior distribution of 
$J^{\rm max}_{\rm CP}$ is perfectly Gaussian and agrees with the profile
likelihood. For $J_{\rm CP}$ large differences appear between 
the posterior distribution and the profile likelihood and lead to 
some difference in the corresponding CL intervals. In particular we find that 
negative values of $J_{\rm CP}$  are preferred in both analysis  but 
more strongly in the Bayesian than in the $\chi^2$ analysis.  

The possible quantification of the correlation between $\theta_{23}$
and $\dcp$ taking into account their circular nature has been
discussed in Sec.~\ref{sec:corre}.  In particular, we have introduced a
new correlation coefficient, $\rho_I$, defined in terms of the
mutual information, which is independent of any boundary
conditions on the variables and is invariant under arbitrary
univariate redefinitions of them.  Quantitatively we always find
stronger correlation between $\dcp$ and $\theta_{23}$ in NO than in
IO.

Finally, we note that a Bayesian analysis is particularly
suited for comparing how much better one model describes the
data compared to another model, a comparison which is quantified in terms
of the Bayes factor of the two models (assuming both models
to be equally probable a priori).  We have applied this to the comparison
between the mass orderings, the octant of $\theta_{23}$, and to 
the presence of CP violation with the following conclusions:
\begin{itemize}
\item  In what regards the comparison between both orderings,
we find that, assuming the same prior probability for both,  
their posterior  probabilities are also very similar:
0.55 for IO and 0.45 for NO with a logarithm of Bayes factor of $-0.2$,
which implies that slight preference for 
inverted ordering is not statistically meaningful. 
\item  Applied to the preference for the octant of $\theta_{23}$ we find 
that the second octant is weakly preferred over the first for the inverted
ordering, but not in the normal nor in the case of 
no assumption or knowledge on the ordering. 
Also due to the relatively bad predictivity of the assumption 
of non-maximal mixing, maximal
mixing is weakly preferred over non-maximal in all orderings. 
\item As for CP violation we find that  although technically CP-violation 
is preferred over CP conservation (either for $\dcp=0$ or
$\dcp=\pi$), the corresponding value of the logarithm of the Bayes factor is
always smaller than 1 in absolute value, i.e., the corresponding 
evidence is not even weak.
\end{itemize}

\section*{Acknowledgments}
This work is supported by Spanish
MINECO grants FPA2012-31880, FPA2012-34694 and FPA2013-46570, by the
Severo Ochoa program SEV-2012-0249 and consolider-ingenio 2010 grant
CSD-2008-0037, by CUR Generalitat de Catalunya grant 2009SGR502, by
USA-NSF grant PHY-09-69739 and PHY-13-16617, and by EU grant FP7 ITN
INVISIBLES (Marie Curie Actions PITN-GA-2011-289442).


\clearpage

\bibliography{biblio}
\bibliographystyle{JHEP}


\end{document}